\documentclass[fleqn,usenatbib]{mnras}

\usepackage{newtxtext,newtxmath}
\usepackage{xspace}

\usepackage[T1]{fontenc}

\DeclareRobustCommand{\VAN}[3]{#2}
\let\VANthebibliography\thebibliography
\def\thebibliography{\DeclareRobustCommand{\VAN}[3]{##3}\VANthebibliography}

\usepackage{graphicx}	
\usepackage{subcaption}
\usepackage{amsmath}	
\usepackage{orcidlink}

\newcommand{\FRB}{FRB~20220912A\xspace}
\newcommand{\nancay}{Nan\c{c}ay\xspace}

\title[Burst properties of hyperactive \FRB]{A \nancay Radio Telescope study of the hyperactive repeating \FRB}

\author[D. C. Konijn et al.]{
David~C.~Konijn \orcidlink{0009-0000-7084-9956},$^{1,2,3}$\thanks{E-mail: konijn@astron.nl}
Dant\'{e}~M.~Hewitt \orcidlink{0000-0002-5794-2360
},$^{1}$
Jason~W.~T.~Hessels \orcidlink{0000-0003-2317-1446},$^{1,2,4,5}$
Isma\"{e}l~Cognard \orcidlink{0000-0002-1775-9692},$^{6,7}$
\newauthor
~Jeff~Huang \orcidlink{0000-0002-8043-0048},$^{1}$
Omar~S.~Ould-Boukattine \orcidlink{0000-0001-9381-8466}, $^{1,2}$
Pragya~Chawla \orcidlink{0000-0002-3426-7606},$^{2}$
Kenzie~Nimmo \orcidlink{0000-0003-0510-0740},$^{8}$ 
\newauthor
~Mark~P.~Snelders \orcidlink{0000-0001-6170-2282},$^{2,1}$
Akshatha~Gopinath \orcidlink{0000-0002-1836-0771},$^{1}$
and Ninisha Manaswini 
\orcidlink{0009-0005-1319-9586}
$^{9}$
\\\\
$^{1}$Anton Pannekoek Institute for Astronomy, University of Amsterdam, Science Park 904, 1098 XH, Amsterdam, The Netherlands\\
$^{2}$ASTRON, Netherlands Institute for Radio Astronomy, Oude Hoogeveensedijk 4, 7991 PD Dwingeloo, The Netherlands\\
$^{3}$Kapteyn Astronomical Institute, University of Groningen, Kapteynborg 5419, 9747 AD, Groningen, The Netherlands\\ 
$^{4}$Trottier Space Institute, McGill University, 3550 rue University, Montr\'eal, QC H3A~2A7, Canada\\
$^{5}$Department of Physics, McGill University, 3600 rue University, Montr\'eal, QC H3A~2T8, Canada\\
$^{6}$Station de Radioastronomie de Nan\c{c}ay, Observatoire de Paris, PSL University, CNRS, Universit\'{e} d’Orl\'{e}ans, F-18330 Nan\c{c}ay, France\\
$^{7}$Laboratoire de Physique et Chimie de l’Environnement et de l’Espace LPC2E UMR7328, Universit\'{e} d’Orl\'{e}ans, CNRS, F-45071 Orl\'{e}ans, France\\
$^{8}$MIT Kavli Institute for Astrophysics and Space Research, Massachusetts Institute of Technology, 77 Massachusetts Ave, Cambridge, MA 02139, USA\\
$^{9}$Max Planck Institute for Radio Astronomy, University of Bonn, Auf dem Hügel 69, D-53121, Bonn, Germany\\
}

\date{Accepted XXX. Received YYY; in original form ZZZ}

\pubyear{2024}

\begin{document}
\label{firstpage}
\pagerange{\pageref{firstpage}--\pageref{lastpage}}
\maketitle

\begin{abstract}
The repeating fast radio burst source \FRB was remarkably active in the weeks after its discovery. Here we report 696 bursts detected with the \nancay Radio Telescope (NRT) as part of the Extragalactic Coherent Light from Astrophysical Transients (\'ECLAT) monitoring campaign. We present 68 observations, conducted from October 2022 to April 2023, with a total duration of 61\,hours and an event rate peaking at $75^{+10}_{-9}$\,bursts per hour above a fluence threshold of 0.59\,Jy\,ms in the $1.2-1.7$-GHz band. Most bursts in the sample occur towards the bottom of the observing band. They follow a bimodal wait-time distribution, with peaks at 33.4\,ms and 67.0\,s. We find a roughly constant dispersion measure (DM) over time ($\delta$DM\,$\lesssim$\,2\,pc\,cm$^{-3}$) when taking into account `sad-trombone' drift, with a mean drift rate of $-8.8\,$MHz\,ms$^{-1}$. Nonetheless, we confirm small $\sim0.3$\,pc\,cm$^{-3}$ DM variations using microshot structure, while finding that microstructure is rare in our sample -- despite the 16\,$\upmu$s time resolution of the data. The cumulative spectral energy distribution shows more high-energy bursts ($E_\nu \gtrsim 10^{31}$\,erg/Hz) than would be expected from a simple power-law distribution. The burst rate per observation appears Poissonian, but the full set of observations is better modelled by a Weibull distribution, showing clustering. We discuss the various observational similarities that \FRB shares with other (hyper)active repeaters, which as a group are beginning to show a common set of phenomenological traits that provide multiple useful dimensions for their quantitative comparison and modelling.
\end{abstract}




\section{Introduction}
Fast radio bursts (FRBs) are short-duration, extremely luminous pulses of coherent radio emission originating from unknown extragalactic sources \citep[][]{lorimer2007bright, thornton2013population, petroff2022fast}. FRBs typically have durations ranging from microseconds \citep{snelders2023detection} to milliseconds, and have been observed at radio frequencies ranging from 110\,MHz \citep{pleunis2021lofar} to 8\,GHz \citep{gajjar2018highest}. Thousands of sources of FRBs have been detected, with $\sim$97 per cent observed as (apparently) one-off events \citep{andersen2023chime}. The remaining small fraction of FRBs are known to repeat \citep[e.g.,][]{spitler2016repeating}, often in a clustered manner \citep[e.g.,][]{li2019statistical, hewitt2022arecibo} and showing band-limited emission \citep{gourdji2019sample}.

The bursts from these repeaters are, on average, longer in duration and narrower in bandwidth than those seen from apparently non-repeating FRBs \citep{pleunis2021fast}. Despite these spectro-temporal differences between the repeaters and non-repeaters, some factors point towards a single population of FRBs with a continuous distribution of repetition rates that are capable of producing bursts with different (observed) morphologies. Most importantly, the upper limits on repetition for the apparent non-repeating FRBs are not clearly distinct from the repetition rates observed in the vast majority of repeaters, though with with some notable exceptions \citep{andersen2023chime}. To then consolidate the implied relationship between repetition rate and burst duration, beaming could be invoked to explain temporal differences \citep{connor2020beaming, kumar2024origins}. Most notably, the high end of the burst energy distribution from the repeating FRB~20201124A resembles that of the non-repeating FRB population \citep{kirsten2024link}.

The source(s) and emission mechanism(s) behind repeating and non-repeating FRBs remain unknown, but given the diversity observed in FRB host environments, burst properties and repetition rates, attributing all observed FRBs to a single progenitor is likely an oversimplification. Nevertheless, among the various models, magnetically powered neutron stars, commonly referred to as `magnetars', are among the most promising candidates \citep[e.g.,][]{margalit2019fast, gourdji2020constraining}. The energetics \citep{margalit2020constraints}, timescales \citep{camilo2008magnetar,nimmo2021highly,nimmo2022phasespace,hewitt2023dense}, and emission frequencies \citep[e.g.,][]{dai2019wideband} of magnetars and FRBs suggest a connection between them. The most important observational connection to magnetar theories is the high-luminosity radio burst that was detected from the Galactic magnetar SGR~1935+2154, which showed similarities to FRBs in terms of its duration, spectrum and luminosity \citep{bochenek2020fast, chime2020bright, kirsten2021detection}. However, the rate of SGR~1935+2154-like bursts alone is insufficient to account for the observed FRB rate and repetition fraction \citep{margalit2020implications}. These authors argue that an additional population of magnetars is required, with stronger magnetic fields than Galactic magnetars, but lower birth rates. Indeed, the discovery of an FRB source in a globular cluster \citep{kirsten2022repeating}, further advocates for diverse formation channels, assuming magnetars as the progenitors of FRBs.

On 2022 September 12, \FRB was discovered by the CHIME/FRB Collaboration \citep{mckinven2022nine}. In only three days, nine bursts were detected, indicating an exceptionally high burst rate --- especially considering that, after six years of monitoring by CHIME/FRB, the median number of bursts detected per repeater is three \citep{andersen2023chime}. Shortly after the discovery of \FRB, the source was localised by the Deep Synoptic Array (DSA-110) to its host galaxy PSO~J347.2702+48.7066 at a redshift of $z=0.0771$ \citep{ravi2023deep}. Thereafter, the European VLBI Network (EVN) localised the source to milliarcsecond precision, with coordinates RA (J2000) = $23^{\rm h}09^{\rm m}04.8989^{\rm s}\pm5\,$\rm{mas} Dec (J2000) = $48^\circ42^\prime23.9078^{\prime\prime}\pm5\,$\rm{mas} \citep{hewitt2023milliarcsecond}. These observations also ruled out the presence of a compact persistent radio source (PRS), as has been associated with some other active repeaters \citep{marcote2020repeating,niu2022repeating, bhandari2023constraints}. Multiple telescopes have detected more than a hundred bursts within a few hours of observation around $\sim$\,1.4\,GHz \citep[e.g.,][]{feng2022extreme, kirsten2022precise,zhang2022fast}. Remarkably, at the peak of its activity, the burst rate of \FRB (above a fluence threshold of 100\,Jy\,ms) constituted up to a few per cent of the all-sky rate of FRBs in general \citep{sheikh2022bright}. This rate of high-fluence bursts is comparable to another hyperactive repeater, FRB~20201124A \citep{kirsten2024link}. Such high activity is short-lived, however,  with considerable daily fluctuations \citep{zhang2023fast}. In terms of spectral range, in addition to the detections at 1.4\,GHz \citep{pelliciari2024northern}, \FRB has been detected at $300-400$\,MHz \citep{bhusare2022ugmrt, ould2022bright}, and around 2\,GHz \citep{perera2022detection, rajwade2022detection}, yet no bursts have been found in searches above 3\,GHz or below 200\,MHz \citep{kirsten2022precise, rajwade2022detection, sheikh2022bright}.

The activity of repeaters is often clustered in time \citep{oppermann2018non} and triggered by an unknown mechanism \citep[peharps starquakes or magnetar flares; e.g.,][]{wang2018frb}, as there have been numerous reported `burst storms' where a repeating FRB produces bursts at a rate of dozens to hundreds per hour \citep[e.g.,][]{li2021bimodal, nimmo2023burst, jahns2023frb}. \FRB is the second FRB, after FRB~20201124A \citep{lanman2022sudden,xu2022fast}, that has seemingly awoken from quiesence before entering a hyperactive state during which it exhibits prolonged activity for weeks to months, observable by numerous telescopes in multiple frequency bands. More recently, FRB~20240114A has been seen to do the same \citep{shin2024chime}. CHIME/FRB monitored these sources for years without detecting any bursts, before their spontanteous flare ups. Roughly speaking, the discovery rate of hyperactive FRBs is currently on the order of $\sim$1 per year. It is not known whether these intense periods of activity signal an awakening from quiescence, or the birth of a new source, but studying the evolution of the burst properties during these periods can inform theory. 

In this paper, we report on 696 bursts that have been detected from the hyperactive repeating \FRB, using the \nancay Radio Telescope (NRT) as part of the Extragalactic Coherent Light from Astrophysical Transients (\'ECLAT) monitoring campaign. Three of these bursts were the subject of in-depth analysis in \cite{hewitt2023dense}, which revealed that some bursts from \FRB feature extremely bright, broadband, clustered shots of emission with durations on the order of microseconds. These shots are potentially produced by a different emission mechanism than the broader burst components that show substantial drift not related to dispersion -- i.e., the time-frequency drift \citep{hessels2019frb} that is colloquially referred to as the `sad-trombone' effect. Here, we have analysed the spectro-temporal properties and the energetics of all 696 bursts in our NRT data, drawing population-wide statistics for comparison with other FRB sources. Section~\ref{sec:Obs} outlines the observational setup and data acquisition, followed by a description of how we obtained the burst sample in Section~\ref{sec:SaD}. The burst property analyses are explained in Section~\ref{sec:BA}. We discuss the results in Section~\ref{sec:Disc}, in the context of other \FRB observations, and finally present our main conclusions in Section~\ref{sec:Concl}.

\section{Observations}\label{sec:Obs}
We observed \FRB as part of the \'ECLAT (PI: D.~M.~Hewitt) FRB-monitoring campaign on the NRT. The NRT is a Kraus-type transit telescope located in \nancay, France; it is operated by the {\it Observatoire de Paris}, and associated with the French {\it Centre National de la Recherche Scientifique} (CNRS). At 1.4\,GHz, the NRT has a system temperature of $T_{\text{sys}} \sim 35$\,K and a gain of $G \sim 1.4$\,K\,Jy$^{-1}$, making it effectively as sensitive as a radio dish with a diameter of 94\,m. \'ECLAT was established at the start of 2022 and has since been monitoring about a dozen FRB sources with approximately weekly cadence to constrain their activity levels and map the evolution of burst properties. Between 2022 October 15 and 2023 April 20, we observed \FRB at a cadence of $\sim$2.5 times per week. There were 68 observations in total, each lasting $\sim$1 hour, as shown in Table~\ref{tab:obs_log}. 

Our observations used the NRT's Low-Frequency receiver ($1.1-1.8$\,GHz total span) and the \nancay Ultimate Pulsar Processing Instrument \citep[NUPPI;][]{desvignes2011new}. We recorded data at a central observing frequency of 1.484\,GHz, with 512\,MHz of bandwidth (from $1.228-1.740$\,GHz) split into 128 4-MHz\,channels, with 16 channels from each of the 8 NUPPI nodes. NUPPI provided a time resolution of 16\,$\upmu$s and full Stokes information in a linear basis. For pointing, we used the original DSA-110 localisation of \FRB: RA (J2000) = $23^{\rm h}09^{\rm m}05.49^{\rm s}$ and Dec (J2000) = $48^\circ42^\prime25.6^{\prime\prime}$ \citep{ravi2023deep}. This position is 5.8$''$ offset from the EVN localisation \citep{hewitt2023milliarcsecond}, but the true position is still well within the half-power beam width of the NRT, which is at least a few arcminutes at 1.4\,GHz. 

To maintain the best possible time resolution, our observations were coherently dedispersed within each 4-MHz channel, to a DM of $219.46$\,pc\,cm$^{-3}$ \citep{mckinven2022nine}. The maximum DM offset at the bottom of the observing band that corresponds to a temporal smearing equal to the time resolution of the NRT data (16\,$\upmu$s) is 0.9\,pc\,cm$^{-3}$. In other words, any burst with a DM lower than $218.56$\,pc\,cm$^{-3}$ or higher than $220.36$\,pc\,cm$^{-3}$ will experience residual temporal smearing due to imperfect coherent dedispersion. Since the best-fit DM falls within this range, we can be confident that the bursts do not experience significant intra-channel DM smearing.

\section{Search and Discovery of Bursts}\label{sec:SaD}
In this work, we have expanded on the \'ECLAT pipeline presented by \citet{hewitt2023dense}, to find bursts from \FRB. In short, the pipeline combines the coherence products AA and BB from the eight recorded subbands to create Stokes I filterbanks with the full 512-MHz bandwidth, before flagging radio frequency interference (RFI), in frequency but not time, using the \texttt{rfifind} tool from the \texttt{PRESTO} pulsar software suite \citep{ransom2011presto}. Thereafter, a boxcar match-filter search is conducted for impulsive transient signals, between a DM of 200 and 250\,pc\,cm$^{-3}$, using the GPU-accelerated software \texttt{HEIMDALL}\footnote{\texttt{https://sourceforge.net/projects/heimdall-astro/}}. \texttt{Heimdall} searches the full 512-MHz band, using a S/N threshold that we set at seven. The resulting burst candidates are classified as being either astrophysical or RFI using the deep-learning models `A' to `H' from the machine-learning classifier \texttt{FETCH} \citep[Fast Extragalactic Transient Candidate Hunter, ][]{agarwal2020fetch, agarwal_aggarwal_2020}. Finally, all candidates assigned a $>0.5$ probability of being astrophysical are manually inspected to confirm they are from \FRB. 

In our pipeline, \texttt{FETCH} frequently assigned a $>0.5$ probability of being an astrophysical transient to RFI and, less often, FRBs were misclassified as RFI. \citet{DavidThesis} established that within a sample of 7100 \texttt{Heimdall} candidates containing 263 FRBs from \FRB, \texttt{FETCH} incorrectly classified 54\, FRBs (though note that \texttt{FETCH} has not been retrained specifically on the NRT data). To mitigate this, we incorporated \texttt{FETCH} into a new classification method which we term \texttt{CATCH} \citep[Classification Algorithm and Transient Candidate Handler;][]{DavidThesis}. \texttt{CATCH} classifies FRB candidates based on the structure in the DM-time space. A detailed description of \texttt{CATCH} is presented in Appendix~\ref{sec:catch}. 

\begin{figure*}
    \centering
    \includegraphics[width=0.99\linewidth]{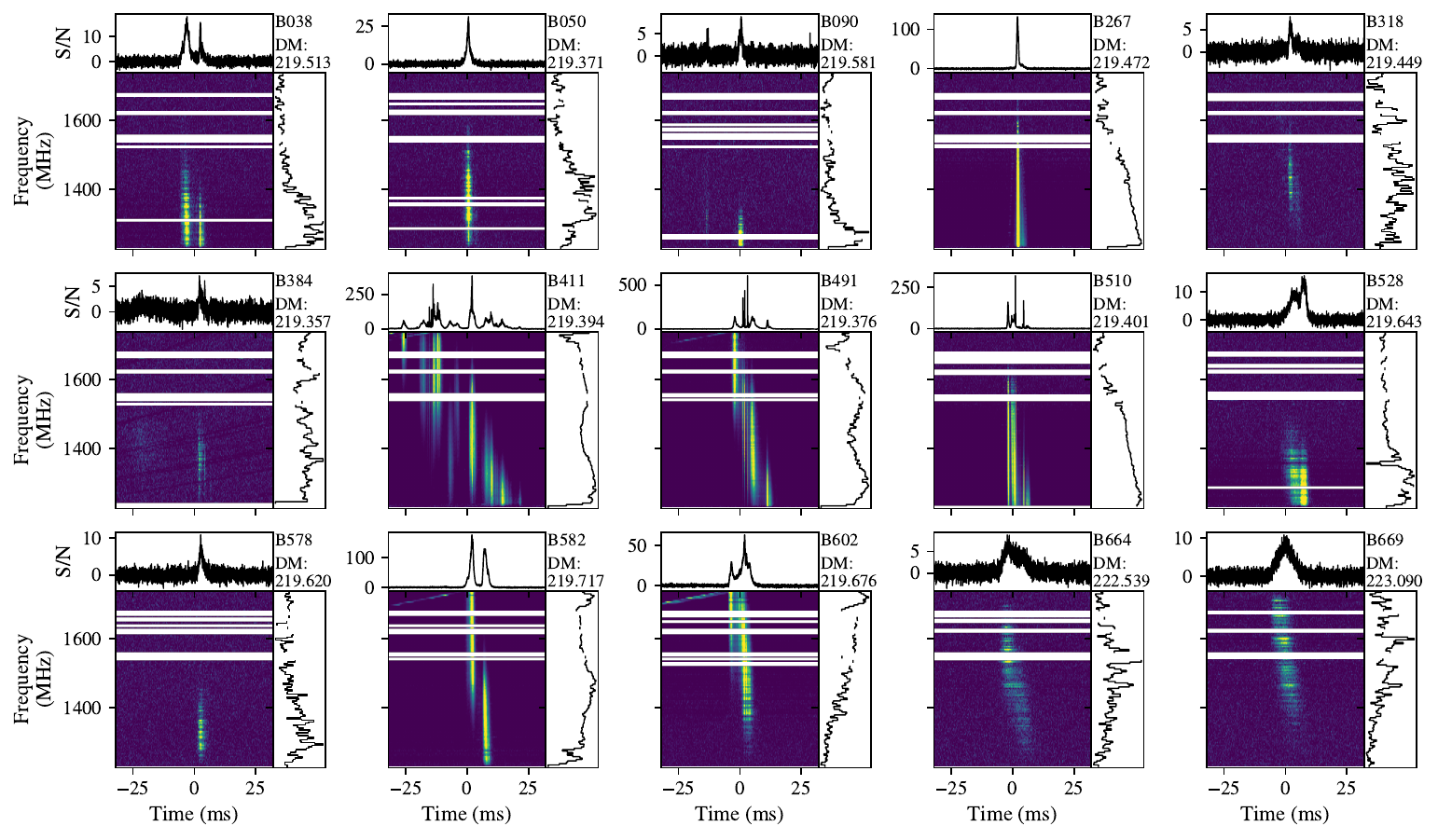}
    \caption{A subset of the 696 NRT-detected bursts from \FRB. This sample consists of very bright bursts, some of which show multiple sub-bursts and/or microsecond-time-scale intensity fluctuations. Each thumbnail shows a dynamic spectrum (bottom-left panel; $\Delta t=16\,\upmu$s and $\Delta \nu=4\,$MHz), frequency-integrated lightcurve (top panel; integrated over the entire frequency band), and time-integrated spectrum (bottom right panel; integrated over the entire time range). All bursts have been coherently dedispersed within each channel to DM\,=\,219.46\,pc\,cm$^{-3}$ during recording. Additionally, each burst has been incoherently dedispersed between channels to the DM value (units pc\,cm$^{-3}$) reported in the top-right of every thumbnail. Also noted in the top-right of each thumbnail is the burst ID, which corresponds to the entries in Tables~\ref{tab:frbtable} and \ref{tab:dmtable}. Horizontal white lines in the dynamic spectra represent channels removed due to the presence of RFI. For visual purposes, the colour maps of the dynamic spectrum range from the value 0.0 to the 99$^{\rm{th}}$ percentile. The plotted S/N values for these bursts are lower than those indicated in Table~\ref{tab:frbtable} because this data has not been downsampled, and the S/N is calculated across the entire bandwidth. A detailed analysis of bursts B411, B491, and B582 was previously presented in \citet{hewitt2023dense}.}
    \label{fig:family}
\end{figure*}

\texttt{Heimdall} has identified a total of 34546 potential FRB candidates, after which our \texttt{CATCH} classification found 696 FRBs with a subband S/N $>$7.0. Here, the subband S/N is the S/N value calculated only in the exact frequency and time region of the burst. At the start of the observations, the event rate was $\sim$60 burst per hour, which decreased and stabilised near the end of the observations to $\sim$1.3 bursts per hour. A subset of the bursts is shown in Figure~\ref{fig:family}. These bursts are not representative of the complete set, but were selected to showcase bursts with high S/N and detailed structures.

\section{Burst properties and analysis} \label{sec:BA}
Table~\ref{tab:frbtable} summarises the burst properties. In the following sub-sections we describe how these properties were measured. While the \'ECLAT observations include full polarimetric information, we differ the polarimetric analysis of these data to a future paper.

\begin{table*}
\caption{\,~ Properties of our sample of 696 NRT-detected bursts from \FRB (full table available online).}
\label{tab:frbtable}
\begin{tabular}{l c c c c c c c}
\hline \hline
Burst ID   & Time of arrival$^{\text{a}}$   & Detection DM$^{\text{b}}$ & Temporal width$^{\text{c}}$ & Frequency extent$^{\text{c}}$  & Central observed frequency$^{\text{c}}$   & S/N$^{\text{c, d}}$            & Fluence$^{\text{c}}$   \\ 
           & (MJD)                          &(pc\,cm$^{-3}$)   & (ms)            & (MHz)        & (GHz)              &                & (Jy ms)     \\ \hline
B000       & 59867.90178241                 & 222.94         & 5.63           & 140          & 1.462              & 26.8         & 3.06    \\
B001       & 59867.90178356                 & 222.94         & 9.12            & 132          & 1.322              & 16.1         & 2.90    \\
B002       & 59867.90209490                 & 226.96         & 5.89           & 120          & 1.286              & 22.1         & 4.02    \\ 
B003       & 59867.90456018                 & 218.94         & 6.66           & 216          & 1.538              & 24.1         & 2.83    \\
B004       & 59867.90456053                 & 218.94         & 8.32            & 200          & 1.398              & 23.2         & 2.83    \\ 
B005       & 59867.90502314                 & 226.96         & 5.38           & 348          & 1.414              & 14.1         & 1.18     \\ 
...         & ...                           & ...               & ...           & ...            & ...               &...               &... \\
\hline
\multicolumn{8}{l}{$^{\text{a}}$ Quoted to $\sim1$\,ms precision. Corrected to the Solar System Barycentre at infinite frequency using the per-burst determined DM, a dispersion constant of}\\
\multicolumn{8}{l}{$^{\text{\ \ }}$ 1/(2.41$\times$10$^4$) MHz$^2$ pc$^{-1}$ cm$^3$ s, and the EVN-derived \FRB position \citep{hewitt2023milliarcsecond}\footnote{\texttt{https://github.com/MSnelders/FRB-Burst-TOA}}. The times quoted are dynamical times (TDB).}\\
\multicolumn{8}{l}{$^{\text{b}}$ As determined by \texttt{Heimdall}.}\\
\multicolumn{8}{l}{$^{\text{c}}$ As defined by \texttt{CATCH} (see Appendix~\ref{sec:catch}).}\\
\multicolumn{8}{l}{$^{\text{d}}$ Subband S/N values calculated on data downsampled by a factor of eight.}\\
\end{tabular}
\end{table*}

\subsection{Spectro-temporal properties}

Our transient classification method, \texttt{CATCH}, determined the frequency extent ($\Delta\nu$) and temporal width ($\tau_{\rm{burst}}$) of each burst by outlining a rectangular box around the burst region in the dynamic spectrum. The positioning of this box was verified manually for each FRB to ensure that the box encompasses all burst emission, but not any high-intensity RFI. We define the middle of the box in frequency extent as the central observed frequency ($\nu_{\rm mid}$) of a burst. 

\begin{figure}
    \centering
    \includegraphics[width=0.95\linewidth]{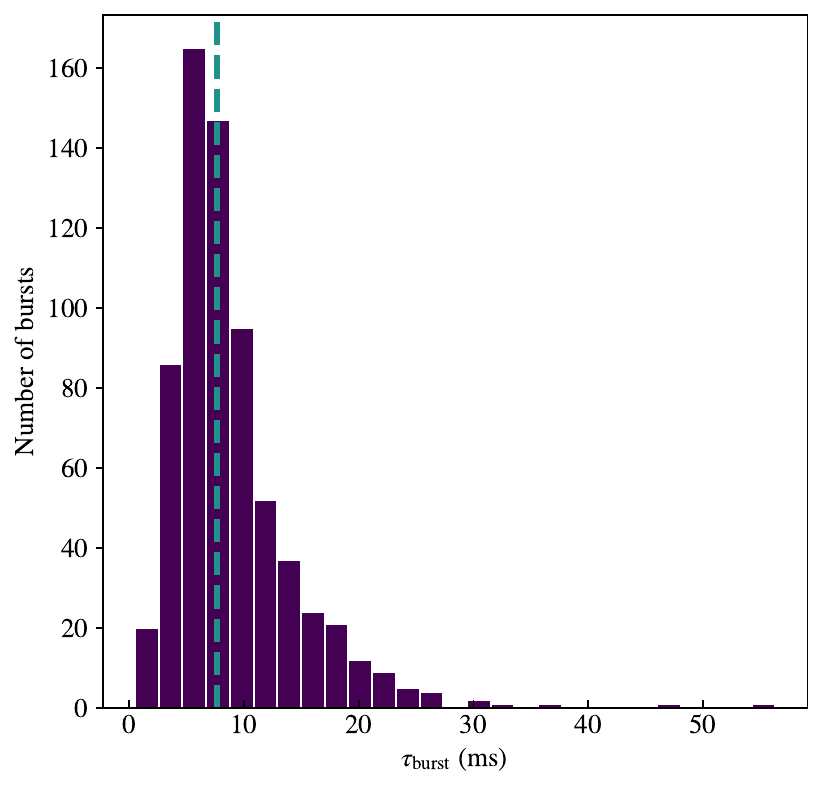}
    \caption{The temporal width distribution for our sample of 696 NRT-detected bursts from \FRB. The vertical dashed blue line indicates the median at 7.68\,ms.}
    \label{fig:temporalwidth}
\end{figure}

Figure~\ref{fig:temporalwidth} shows the temporal width distribution for the 696 bursts we detected from \FRB. Here, the temporal width of the burst is the width of the box that was fitted to the burst, hence it encompasses the earliest and latest times of emission in the case of multi-component bursts. The median temporal width is 7.68\,ms and the distribution is slightly skewed towards longer durations. 

\begin{figure}
    \centering
    \includegraphics[width=1\columnwidth]{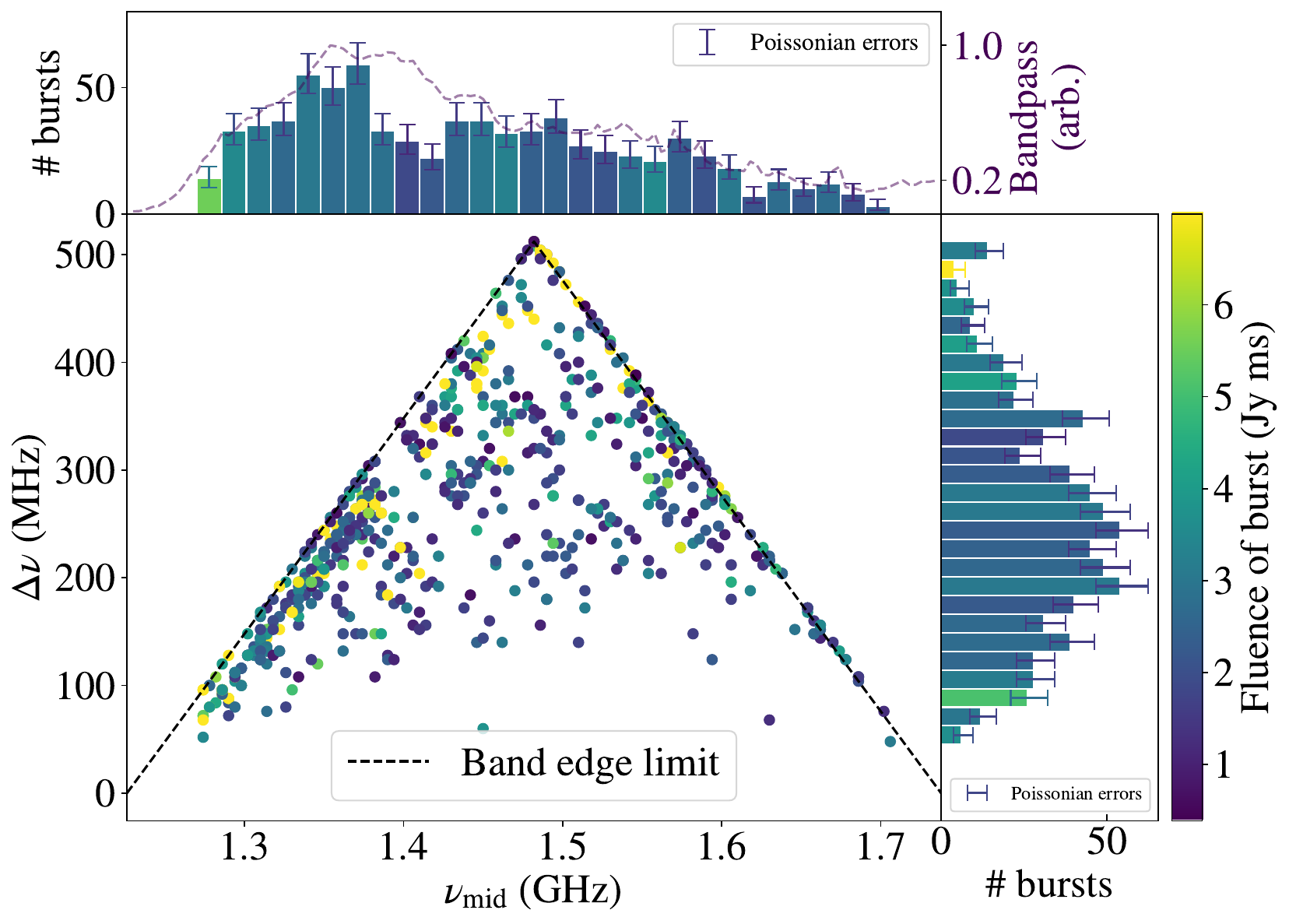}
    \caption{The frequency extent, $\Delta \nu$, of each detected burst from \FRB is plotted against its central observed frequency $\nu_{\rm mid}$, and coloured by its fluence. The right and top panels show the distritbutions of $\Delta\nu$ and $\nu_{\rm mid}$, respectively, where the NRT bandpass has been added to the top histogram. The error bars represent the Poissonian uncertainty, with a 1$\sigma$ confidence interval. The dashed black triangular shape in the middle panel represents the limits of our observing band (see the text for more details). The colour of the bins of the side histograms represents the median fluence of the bursts within each bin.}
    \label{fig:bandwidth_vs_central_freq}
\end{figure}

Figure~\ref{fig:bandwidth_vs_central_freq} shows $\Delta \nu$ against $\nu_{\rm mid}$ for each burst, coloured by fluence. Given that $\nu_{\rm mid}$ represents the middle of the frequency extent where burst emission is visible within our observing band, bursts with larger frequency extents will naturally have $\nu_{\rm mid}$ approaching the center frequency of our observing band. The maximum allowed frequency extent as a function of central frequency is shown by dashed black lines that constrain the 2D distribution to a triangular region. The frequency extent distribution exhibits an artificial peak at the maximum value, which is the result of many bursts having a frequency extent larger than our observing band. The median frequency extent is 248\,MHz, while the central frequency distribution demonstrates a preference for lower central frequencies. The top histogram shows that more bursts are present at lower frequencies. 

\subsection{Fluence and energetics}

The S/N and fluence of a burst are calculated within the spectro-temporal box determined by \texttt{CATCH}. We note that for some bursts the emission clearly extends beyond our observing band, but we refrain from extrapolating and estimating frequency extents and fluences outside of the observing band. In these cases, the quoted values are thus lower limits. We calculated the fluence of a burst, $F$ (Jy\,ms), by integrating the flux with respect to time, using the radiometer equation \citep{lorimer2005handbook}:

\begin{equation}
F = \text{S/N} \cdot \text{SEFD} \cdot \sqrt{\frac{\tau_\text{burst}}{{\rm n}_\text{pol} \Delta \nu}} 
\label{eq:fluence}
\end{equation}

\noindent where SEFD is the system equivalent flux density ($\sim$\,25\,Jy for the NRT) and ${\rm n}_\text{pol}$ is the number of polarisations recorded in the observation (2 in this case).

\begin{figure}
    \centering
    \includegraphics[width=1\linewidth]{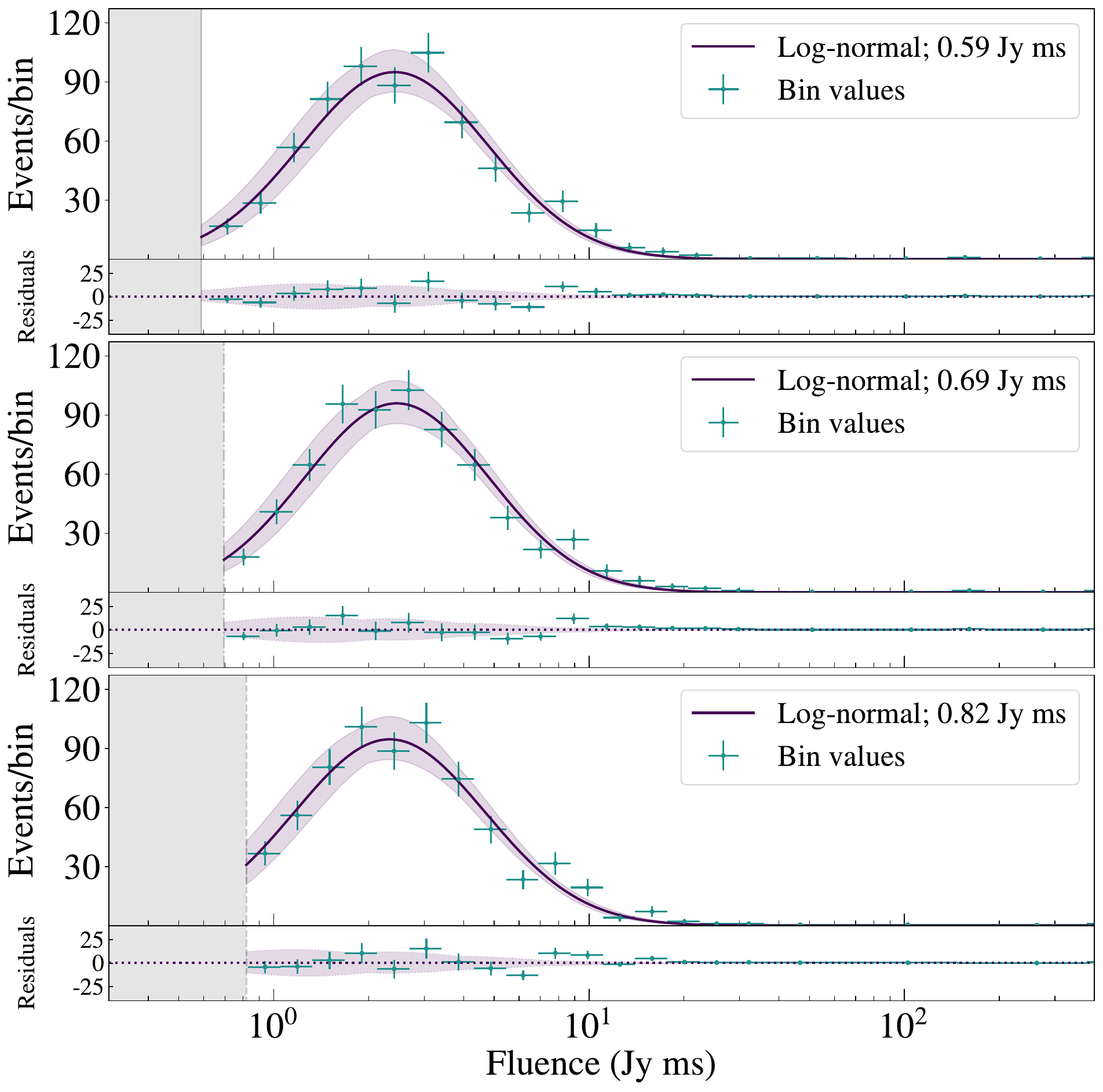}
    \caption{The burst fluence distribution for completeness thresholds 0.59, 0.69, 0.82\,Jy\,ms, calculated for three temporal widths: 5.63, 7.68 and 10.79\,ms. The grey areas indicate the incompleteness regions, whereas the purple areas indicate the 1$\sigma$ fit error of a log-normal function to the distributions. Fit residuals are shown in the smaller panels beneath each fluence distribution.}
    \label{fig:fluences}
\end{figure}
 
Figure~\ref{fig:fluences} shows the fluence distribution of our NRT-detected bursts for the completeness thresholds that depend on assumed burst width. We estimated these completeness thresholds using Equation~\ref{eq:fluence}, assuming a S/N~$=7$, a frequency extent of 248\,MHz (the median value from fitting our burst sample), and three different temporal widths: the first-quartile value from our burst temporal duration distribution (5.63\,ms), the second quartile (or median; 7.68\,ms), and the third quartile (10.79\,ms). We then fit a log-normal distribution to the data above the three aforementioned completeness thresholds. In all cases the reduced $\chi^2$ values (1.31, 1.57 and 1.03) do not exceed the critical $\chi^2$ values (1.58, 1.60 and 1.60), confirming that the fluence distribution is well fit by a log-normal function regardless of the assumed completeness threshold. 

\addtocounter{footnote}{2}
\footnotetext[2]{\texttt{https://github.com/MSnelders/FRB-Burst-TOA}} 

\begin{figure*}
    \centering
    \includegraphics[width=0.95\linewidth]{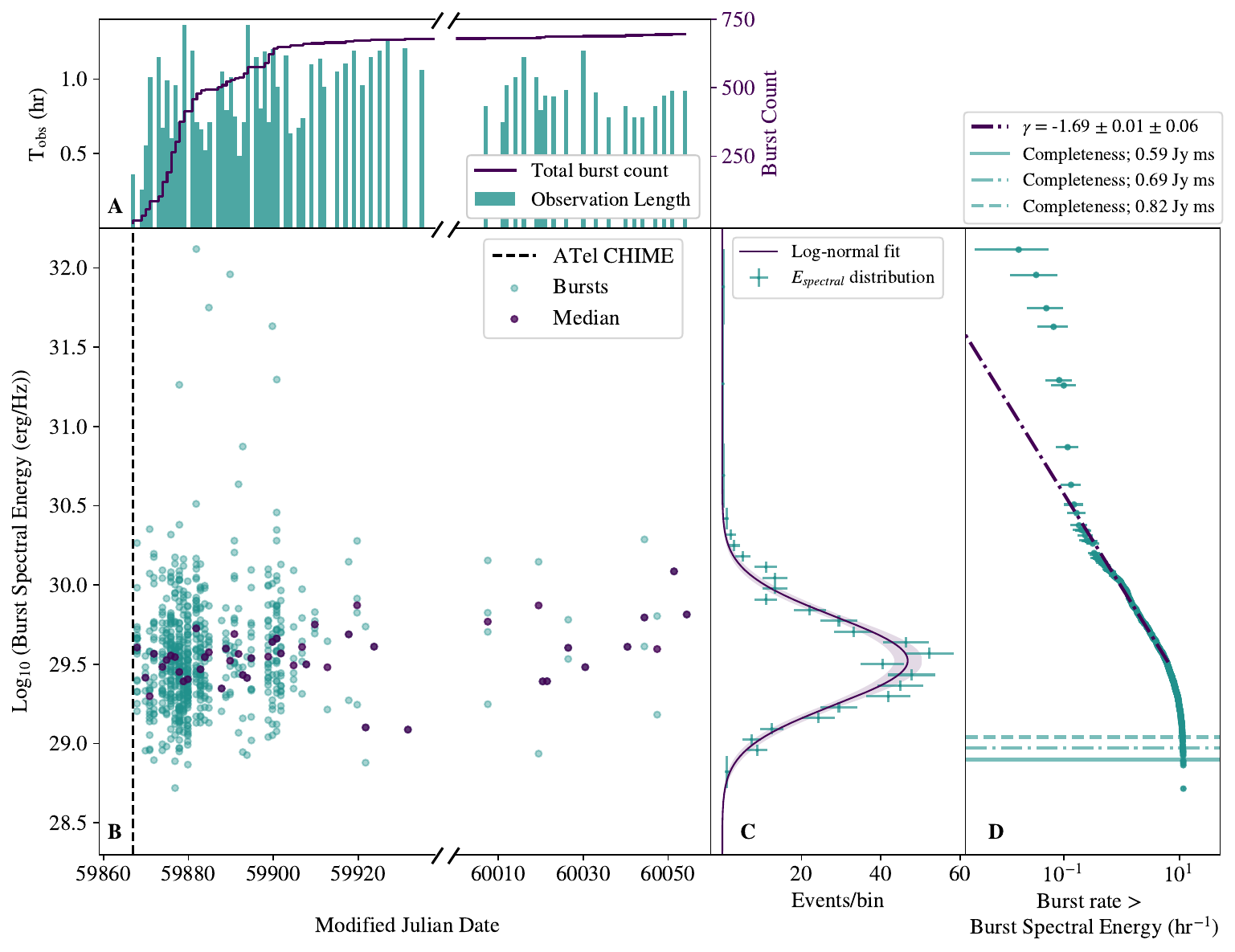}
    \caption{Temporal evolution of the spectral energy distribution of \FRB bursts detected with the NRT. Slashes through Panels A and B indicate a break in the continuity of the plotted timeline. Panel A: The green bars show the duration of each observation, while the solid purple line indicates the cumulative burst count with time. Panel B: The green data points display the spectral energy of each detected burst, while the purple data points show the bursts' median spectral energy on each observing day. The vertical dashed black line indicates the time when CHIME/FRB reported the discovery of \FRB and when we started observing with NRT. Panel C: The spectral energy distribution, considering all the detected bursts, is shown in green with a log-normal fit and the 1$\sigma$ fit error indicted by the purple line and shaded region. The mean lies at $10^{29.48}$\,erg/Hz, with a standard deviation of $10^{0.33}$\,erg/Hz. Panel D: The cumulative spectral energy distribution plotted in green and fitted with a power law, show by the  black dot-dashed line. The horizontal green lines indicate the theoretical completeness thresholds for various assumed burst widths, assuming a minimal S/N$ =7$ detection, as in Figure~\ref{fig:fluences}. The solid green line indicates the completeness threshold for assumed burst widths at 5.63\,ms; the dash-dotted line at 7.68\,ms; and the dashed line at 10.79\,ms. The distribution is seen to flatten at higher energies, indicating that bursts with these energies happen more frequently than predicted by extrapolating a simple power law based on fitting lower-energy bursts. This effect is less obviously present in Panel~C, as several one-burst bins are obscured by the fit line.}
    \label{fig:energydistribution}
\end{figure*}

\newpage
The main panel of Figure~\ref{fig:energydistribution} shows the bursts' spectral energy densities, $E_{\text{spectral}}$, which we calculated as follows \citep[See][Equation~10]{macquart2018frb}:

$$E_{\text{spectral}}= 1\times10^{-23} \cdot \frac{F \cdot 4 \pi D_L^2}{(1+z)^2} \hspace{0.2cm} \text{erg/Hz},$$ 

\noindent where $z$ and $D_L$ are the redshift and luminosity distance to the source \citep[0.0771 and 362.4\,Mpc, respectively;][]{ravi2022host}. 

In this figure, Panel~A shows the duration of each observation (median of 0.92\,hr) together with the cumulative burst count, illustrating the high cadence of our observations. The majority of the bursts that were detected at the beginning of the observational campaign have spectral energy densities in the $10^{29}-10^{30}$\,erg/Hz range. The distribution of spectral energies is shown in Panel~C, where a log-normal function is fitted to the data. The reduced $\chi^2$ value is 1.09, which does not exceed the critical value of 1.39, indicating a good fit. Panel~D shows the cumulative distribution of spectral energies. It depicts the burst rate per hour with spectral energies larger than a specific threshold. The errors on the rate are Poissonian, and we assume a fractional error of 20 per cent for the derived fluence values. The apparent turnover visible at lower energies is a reflection of bursts being detected close to the completeness threshold of the telescope, where it is easier to miss some bursts.

We fit a power law to the distribution using the \texttt{powerlaw} package \citep{clauset2009power}. In addition to performing power-law fits, this package identifies the initial turnover point in the cumulative distribution to determine the appropriate minimum spectral energy from which to start the fit. In the case of a perfectly log-normal spectral energy distribution, the start of the power-law fit (i.e., the mimimum spectral energy being fit) should coincide with the completeness threshold. The three completeness thresholds for bursts with temporal widths of 5.63\,ms 7.68\,ms, and 10.79\,ms deviate from the start of the power law by approximately $10^{0.61}, 10^{0.54},$ and $10^{0.47}$\,erg/Hz, respectively. This could, in principle, indicate that the spectral energy distribution flattens at lower energies. Note that the completeness threshold can also deviate even more from its estimated value, in which case the log-normal distribution would not fit the data well. The fitted power law has a slope of $-1.67 \pm 0.01 \pm 0.06$, where the first error is determined as the quadrature sum of the error on the fit using the fractional 20 per cent fluence errors and the second error is determined by bootstrapping. The data follows the power law until spectral energies of $10^{30.5}$\,erg\,Hz$^{-1}$, after which the distribution appears to flatten. This flattening of the cumulative spectral energy distribution has been seen in other repeaters such as FRB~20121102A \citep{hewitt2022arecibo} and FRB~20201124A \citep{kirsten2024link}, and will be discussed in more detail for \FRB by \citet{R117paper}, using additional data from several other telescopes. 

\begin{figure}
    \centering
    \includegraphics[width=1\columnwidth]{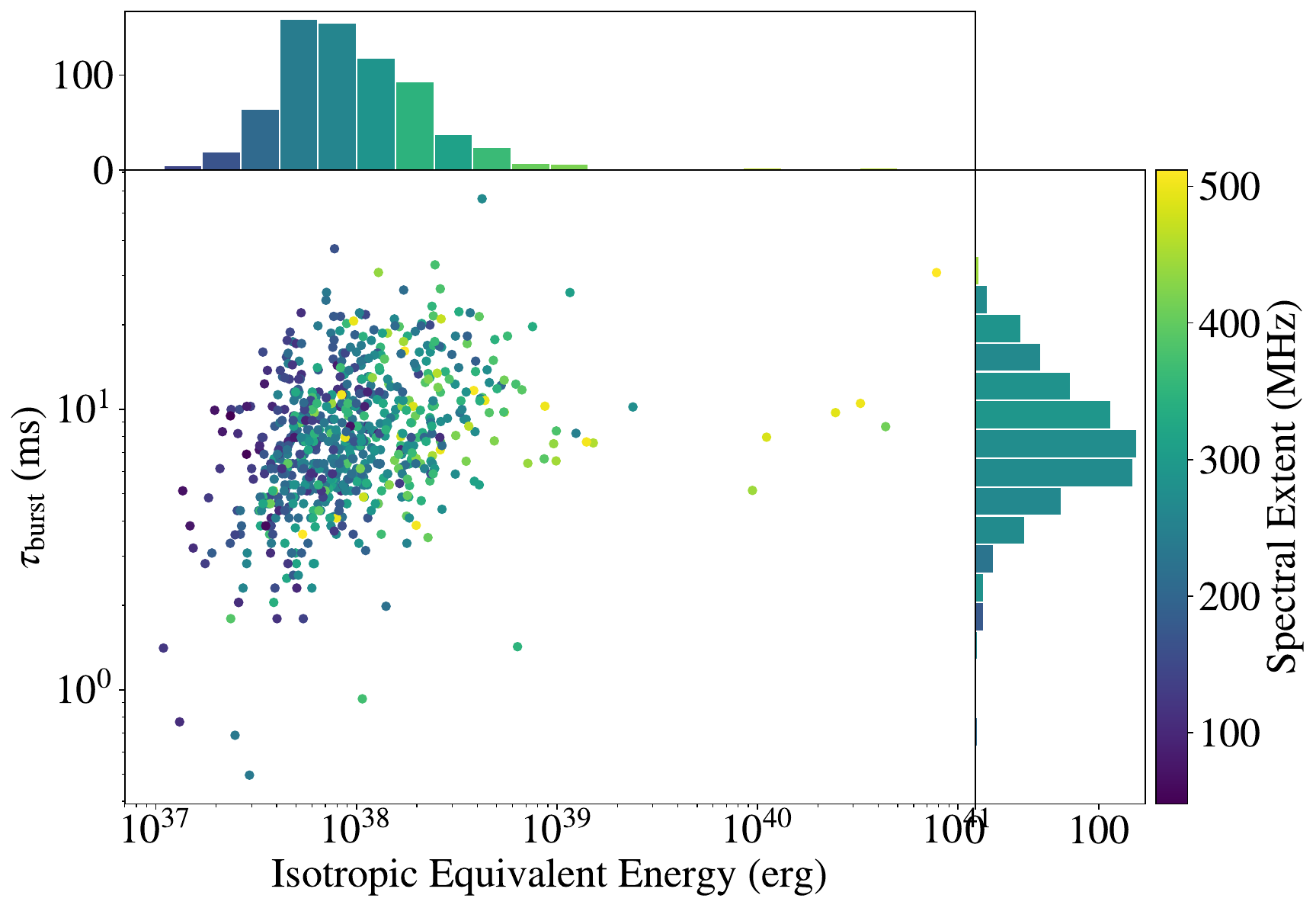}
    \caption{The temporal widths of bursts plotted against their isotropic-equivalent energy. The histograms are coloured based on the median value of the bursts within each bin.}
    \label{fig:TimewidthvsEnergy}
\end{figure}

The isotropic-equivalent energy (in erg) is the total energy released, assuming that the energy is radiated isotropically. We assume isotropic emission due to the unknown beaming angle, despite the expectation that FRB emission is highly directional. The isotropic-equivalent energy can be determined by multiplying the spectral energy by the bandwidth of the burst in Hz. Figure~\ref{fig:TimewidthvsEnergy} shows the temporal widths of the bursts plotted against their isotropic-equivalent energies. The colours of the histograms are based on the median of the bursts inside each bin. Bursts with higher isotropic-equivalent energies and greater temporal widths tend to span larger bandwidths, whereas bursts with lower isotropic-equivalent energies tend to span smaller bandwidths.

\subsection{Dispersion measure and time-frequency drift} \label{Dispersion Measure}
Determining the DM of a burst is non-trivial, due to the complex spectro-temporal morphology seen in many FRBs, along with the lack of {\it a priori} knowledge about how the burst should look. There is no exact model for each individual FRB and effects such as scintilation, the sad-trombone effect \citep{hessels2019frb}, and the residual intra-burst drift not related to dispersion seen in many bursts \citep[e.g.,][]{rajabi2020simple, chamma2021evidence, jahns2023frb} make it difficult and ultimately ambiguous to accurately determine the `true' DM. 
In this paper, we determined the DMs using \texttt{DM\_phase} \citep{seymour2019dm_phase}\footnote[3]{\texttt{https://github.com/danielemichilli/DM\_phase}}, which maximizes the coherent power of a burst across the bandwidth in order to determine a best DM. Using 32-bit, coherently dedispersed (DM\,=\,219.46\,pc\,cm$^{-3}$) filterbank data, we estimated the DM for 15 bursts with subband S/N $>100$ and visible (sub-)millisecond temporal structure, carefully selecting a suitable range of fluctuation frequencies to consider in \texttt{DM\_phase}. The selected bursts are: B038, B050, B090, B267, B318, B384, B411, B491, B510, B528, B578, B582, B602, B664, and B669. Our DM measurements are tabulated in Table~\ref{tab:dmtable}.

\begin{table*}
\caption{\,~ Properties of the \FRB bursts with fluctuation frequencies $>2.5$\,ms$^{-1}$}
\label{tab:dmtable}
\begin{tabular}{l c c c c c c c}
\hline \hline
Burst ID            & Time of arrival$^{\text{a}}$  & DM$^{\text{b}}$       & DM error$^{\text{c}}$          & Min. fluctuation frequency & Max. fluctuation frequency     & Temporal width$^{\text{d}}$    & Spectral extent$^{\text{d}}$ \\ 
                    & (MJD)            & (pc\,cm$^{-3}$)        & (pc\,cm$^{-3}$)    & ($1\times10^{-1}$ ms$^{-1}$)                   & ($1\times10^{-1}$ ms$^{-1}$)                       & (ms)          & (MHz)     \\ \hline
B038                & 59870.88721446   & 219.51               & 0.06             & 60                            & 175         & 8.37         & 380\\
B050                & 59871.87238284   & 219.37               & 0.10             & 50                            & 120         & 1.42         & 348\\
B090                & 59873.88115071   & 219.58               & 0.20             & 10                            & 51          & 5.44         & 128\\ 
B267                & 59877.86661735   & 219.47               & 0.05             & 120                           & 244         & 5.14         & 444\\
B318                & 59879.83220702   & 219.45               & 0.23             & 21                            & 57          & 3.49         & 404\\ 
B384                & 59881.86993869   & 219.36               & 0.01             & 850                           & 2050        & 47.78        & 452\\ 
B411$^{\text{e}}$   & 59881.83351237   & 219.39               & 0.10             & 29                            & 102         & 26.08         & 308\\
B491$^{\text{e}}$   & 59884.83816794   & 219.38               & 0.01             & 1000                          & 3125        & 10.50        & 500\\
B510                & 59889.82448343   & 219.40               & 0.01             & 1000                          & 3125        & 8.67         & 412\\
B528                & 59891.81058320   & 219.64               & 0.10             & 33                            & 112         & 8.21         & 248\\
B578                & 59898.82970864   & 219.62               & 0.22             & 10                            & 50          & 1.98         & 220\\ 
B582$^{\text{e}}$   & 59899.78833990   & 219.72               & 0.02             & 200                           & 550         & 9.73         & 492\\
B602                & 59900.82699925   & 219.68               & 0.06             & 40                            & 220         & 7.95         & 484\\ 
B664                & 59917.75117914   & 222.53               & 0.48             & 7                             & 27          & 6.42         & 440\\ 
B669                & 59919.74566971   & 223.09               & 0.28             & 20                            & 50          & 6.66         & 392\\ 
\hline
\multicolumn{8}{l}{$^{\text{a}}$ Quoted to $\sim1$\,ms precision. Corrected to the Solar System Barycentre at infinite frequency using the per-burst determined DM, a dispersion constant of}\\
\multicolumn{8}{l}{$^{\text{\ \ }}$ 1/(2.41$\times$10$^4$) MHz$^2$ pc$^{-1}$ cm$^3$ s, and the EVN-derived FRB~20220912A position \citep{hewitt2023milliarcsecond}\textcolor{blue}{$^2$}. The times quoted are dynamical times (TDB).}\\
\multicolumn{8}{l}{$^{\text{b}}$ Computed using \texttt{DM\_phase} within the spectro-temporal confines of the burst defined by \texttt{CATCH}.}\\
\multicolumn{8}{l}{$^{\text{c}}$ See Appendix~\ref{ap:dm}.}\\
\multicolumn{8}{l}{$^{\text{d}}$ The temporal width and frequency extent of the burst defined by \texttt{CATCH} (see Appendix~\ref{sec:catch}).}\\
\multicolumn{8}{l}{$^{\text{e}}$ These three bursts are presented in detail in \citet{hewitt2023dense}. The DMs reported here are consistent with the DMs those authors determined, }\\
\multicolumn{8}{l}{$^{\text{\ \ }}$ where they ensured that the extremely short time-scale, broad-band structures (microshots) arrived at all observed frequencies at the same time.}\\
\end{tabular}
\end{table*}

\begin{figure}
    \centering
    \includegraphics[width=1\linewidth]{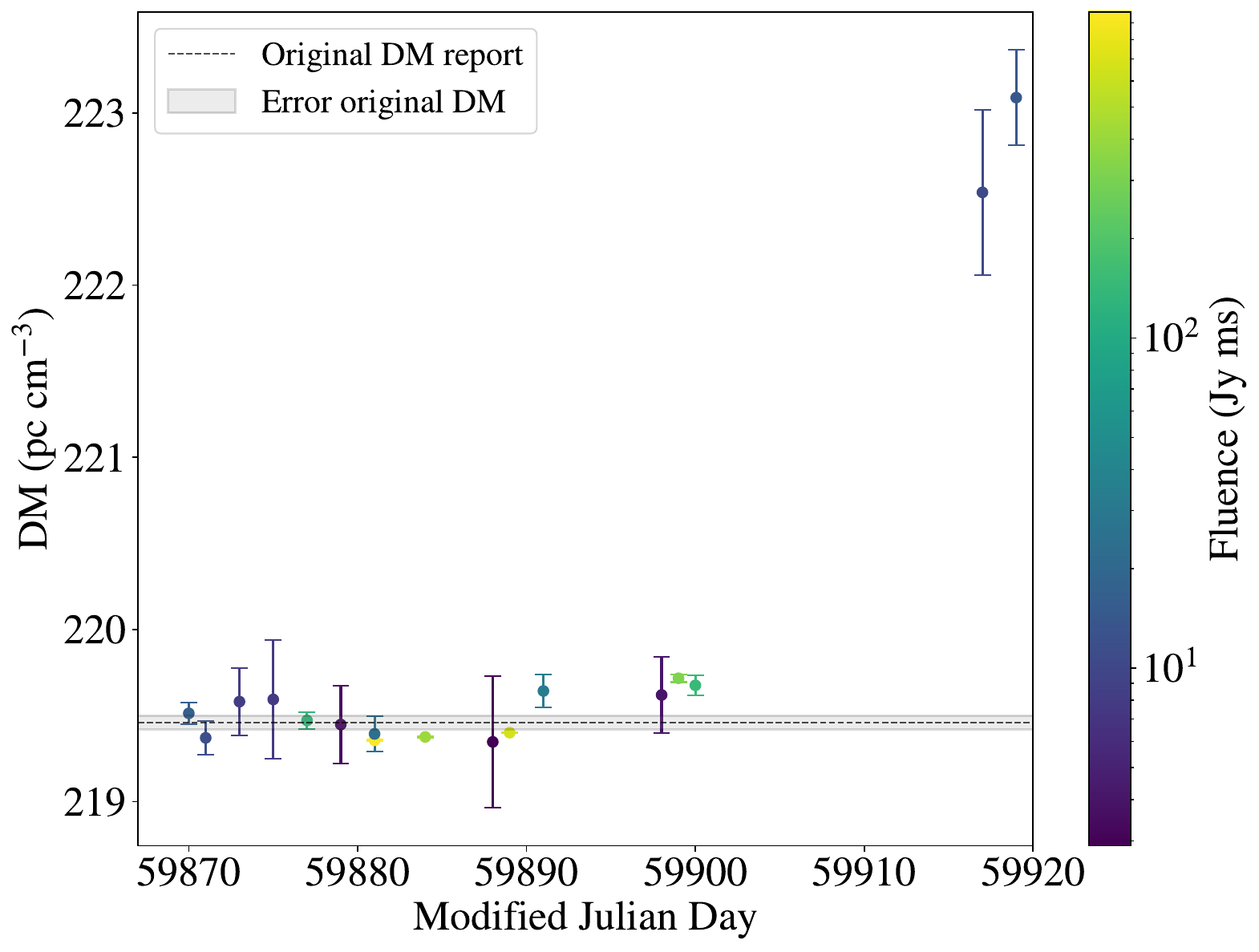}
    \caption{DM measurements for all bursts above a CATCH S/N of 120 and with the maximum fluctuation frequencies exceeding $2.5$\,ms$^{-1}$. This fluctuation frequency limit ensures that the timescales on which the intensity significantly changes are smaller than 4\,ms. The colour of each point indicates the subband S/N of the respective bursts. The dashed horizontal line, along with the shaded region, indicates the original DM and its associated error reported by \citet{mckinven2022nine}. There appears to be a jump in the DM between MJDs 59900 and 59917.}
    \label{fig:DM}
\end{figure}

In Figure~\ref{fig:DM} the DM is plotted as a function of time. The DM appears constant over the majority of our campaign. However, towards the end of our campaign, we observe an apparent increase in the DM of 2.86$\pm$0.54\,pc\,cm$^{-3}$ from Modified Julian Date (MJD) 59900 to MJD~59917. We can determine the validity of this increase in DM by looking at the total drift rate of each bright (S/N$>30$) burst. 

\begin{figure}
    \centering
    \includegraphics[width=1\linewidth]{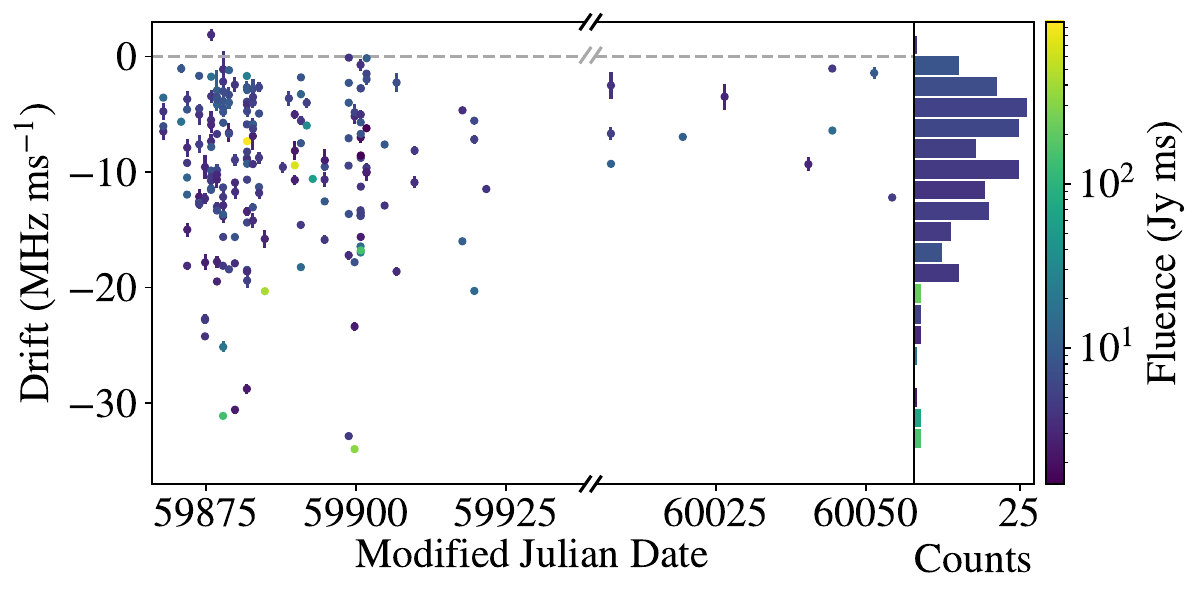}
    \caption{The total drift of each burst with S/N $>30$, where the DM has been fixed at 219.46\,pc\,cm$^{-3}$. The dashed grey line indicates no drift. The histogram bars are coloured based on the median value of the bursts within each bin.}
    \label{fig:drift_bursts}
\end{figure}

The total drift of a burst is mainly determined by three factors: the DM, the sad-trombone-induced drift between sub-bursts, and the intra-(sub-)burst drift. Assuming the influence of the sad-trombone drift and the intra-burst drift remains constant over time, any observed increase in the total drift over time could then be attributed to an increase in the DM. To determine the total drift, we follow the methodology presented in \cite{gopinath2024propagation}, where we fix the DM to 219.46\,pc\,cm$^{-3}$ (the DM used for coherent dedispersion), and measured the burst drift for all bursts that have a S/N$>30$. This entails fitting a 2D Gaussian function to the auto-correlation function (ACF) of the bursts, where the zero-lag noise spike has been masked. The drift rate is then essentially $cot(\theta)$, where $\theta$ is the 
angle of the 2D Gaussian, measured counterclockwise from the positive y-axis to the semi-major axis. The total drift is shown in Figure~\ref{fig:drift_bursts}. 

We find a median drift rate of $-8.8\,$\,MHz\,ms$^{-1}$. The drift rate appears to stay more or less constant throughout our campaign. As the burst rate (apparently) decreases, fewer measurements are available. We performed a Kolmogorov-Smirnov test using \texttt{scipy.stats.kstest}, comparing the drift measurements before and after MJD~60000, to see if there is a statistically significant decline in the magnitude of the drifts over time. We find $p=0.0986$ and $K=0.38$, and thus cannot reject the null hypthesis that the samples are drawn from the same distribution. This lack of an increase in drift, given a fixed DM, also argues for a DM that is (roughly) constant over time. An increase in DM by $\sim2$\,pc\,cm$^{-3}$ would result in significant differences between the samples before and after MJD~60000. It is likely that we have absorbed some of the sad-trombone drift or intra-burst drift in bursts B664 and B669 when determining the DM -- thus leading to an apparently higher DM when naively fitting those bursts.

\subsection{Wait-times} \label{waittimes}

\begin{figure}
    \centering
    \includegraphics[width=1\linewidth]{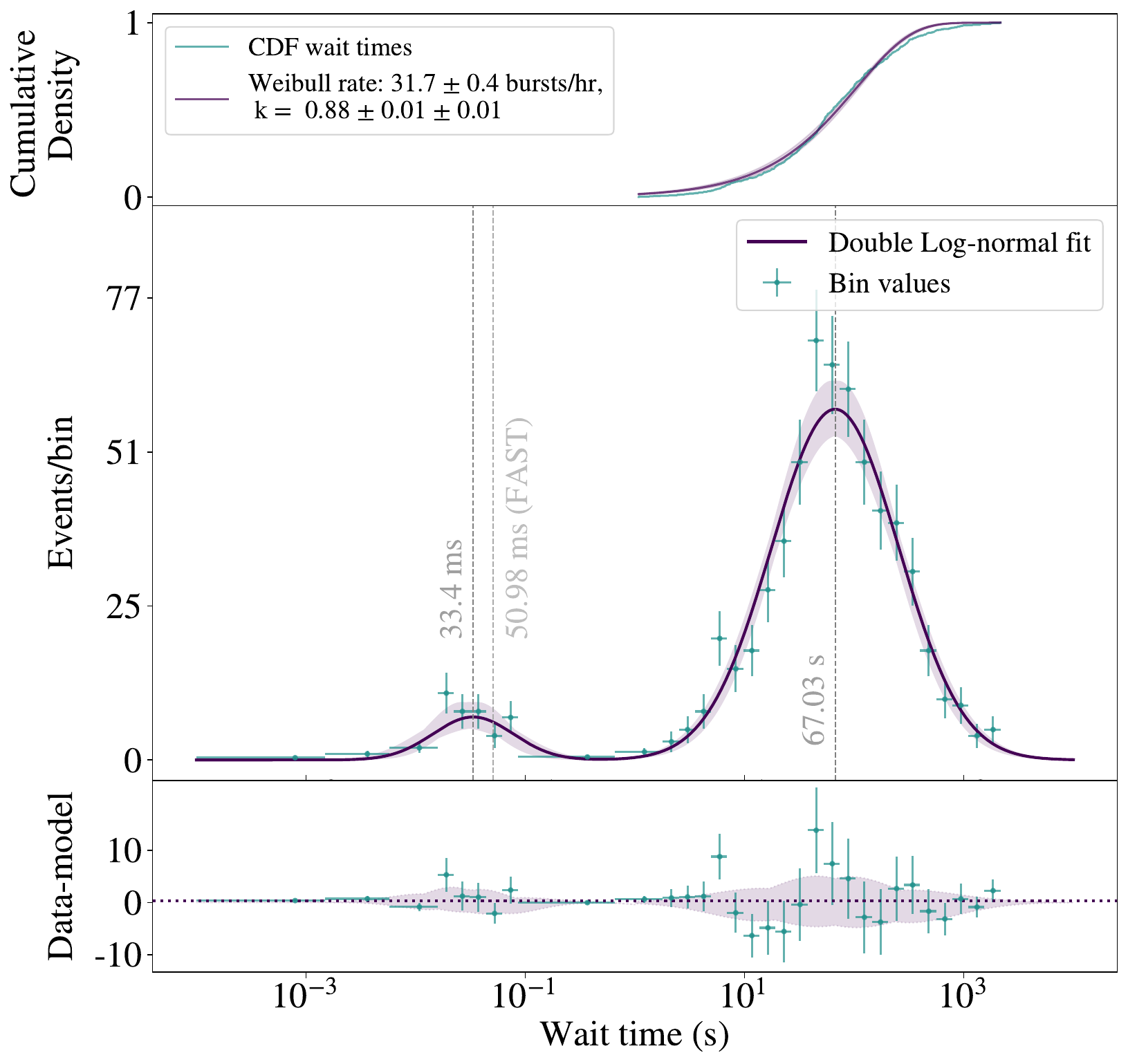}
    \caption{The distribution of the wait-times from \FRB plotted on a logarithmic scale. A double log-normal distribution has been fitted to the data, and the peaks of the wait-time distribution are at $33.40 \pm 6.41$\,ms and $67.03 \pm 2.61$\,s, respectively. Note that the quoted error is the uncertainty on the location of the wait-time peak. The purple areas indicate the 1$\sigma$ fit error and the bottom plot shows the fit residuals. The top plot shows the cumulative wait-time distribution of the second wait-time peak, to which a Weibull cumulative distribution function has been fitted. The total error on the shape parameter $k$ is a combination of the error on the fit, and the error derived from bootstrapping 1000 times. The burst rate deviates from a Poisson distribution, as the shape parameter of the fitted Weibull CDF $k = 0.88\pm0.01\pm0.01$ is not equal to 1.0.}
    \label{fig:waittimes}
\end{figure}

The time between successive bursts is referred to as the `wait-time'. We consider two bursts to be separate, rather than multiple components of a single burst, if the signal between the two high-intensity regions returns to the level of the noise. In Figure~\ref{fig:waittimes} we show the wait-time distribution for our sample of NRT-detected bursts. As for several other repeaters, e.g. FRB~20121102 \citep{li2021bimodal}, FRB~20201124A \citep{xu2022fast}, and FRB~20200120E \citep{nimmo2023burst}, the distribution is bimodal. We fit two log-normal functions to the peaks, yielding a reduced $\chi^2$ value of 1.08, which is below the critical threshold of 1.36, indicating a good fit. To minimise the effect of outlier data points, and to more robustly estimate uncertainties, we have bootstrapped the data by randomly selecting 90 per cent of the data without replacement 1000 times, and only fitting that sample with a log-normal function. From our bootstrapping analysis, we find wait-time peaks at $33.40 \pm 6.41$\,ms and $67.03 \pm 2.61$\,s. Note that the quoted error is the uncertainty on the location of the wait-time peak. The full width at half maximum of the shorter-timescale (left) peak at 33.40\,ms is 63.30\,ms.

Also shown in Figure~\ref{fig:waittimes} is the cumulative wait-time distribution for wait-times corresponding to the second peak at longer times. We can use this cumulative wait-time distribution to test if the burst rate is Poissonian, over the course of our observing campaign, or if there is clustering in time. If the bursts are distributed in time according to a Poisson process, then the distribution of intervals $t_{\text{wait}}$ between subsequent bursts would have an exponential distribution. One generalisation of an exponential distribution of intervals is the Weibull distribution 

\begin{equation}
    \mathcal{W}(t_{\text{wait}}|k, r) = k\,t_{\text{wait}}^{-1} \,[t_{\text{wait}}\, r\, \Gamma (1+1/k)]^k~\text{e}^{-[t_{\text{wait}} \,r \,\Gamma (1+1/k)]^k},
\end{equation}

\noindent where $r$ is the burst rate, $\Gamma$ is the gamma function, and $k$ is the shape parameter \citep{papoulis2002probability}. The cumulative density function (CDF) of the Weibull distribution can be fitted to the cumulative wait-time distribution to test for possible burst clustering \citep[e.g.,][]{oppermann2018non}. The Weibull CDF is defined as:

\begin{equation}
    P_{\text{Weibull}} = 1 - e^{-(t_{\text{wait}}\,r \,\Gamma(1+1/k))^k}.
\end{equation}

This function is identical to the Poisson CDF when $k = 1$, while $k \neq 1$ implies that the bursts have a variable rate and cluster in time.

We find that the best-fit Weibull burst rate is $31.7 \pm 0.4$\,bursts/hour, with a shape parameter of $0.88 \pm 0.01 \pm 0.01$. The total error on the shape parameter is a combination of the error on the fit and the error derived from bootstrapping. As before, the data is bootstrapped 1000 times. 

\subsection{Burst rate} \label{Burst Rate}

\begin{figure*}
    \centering
    \includegraphics[width=0.9 \linewidth]{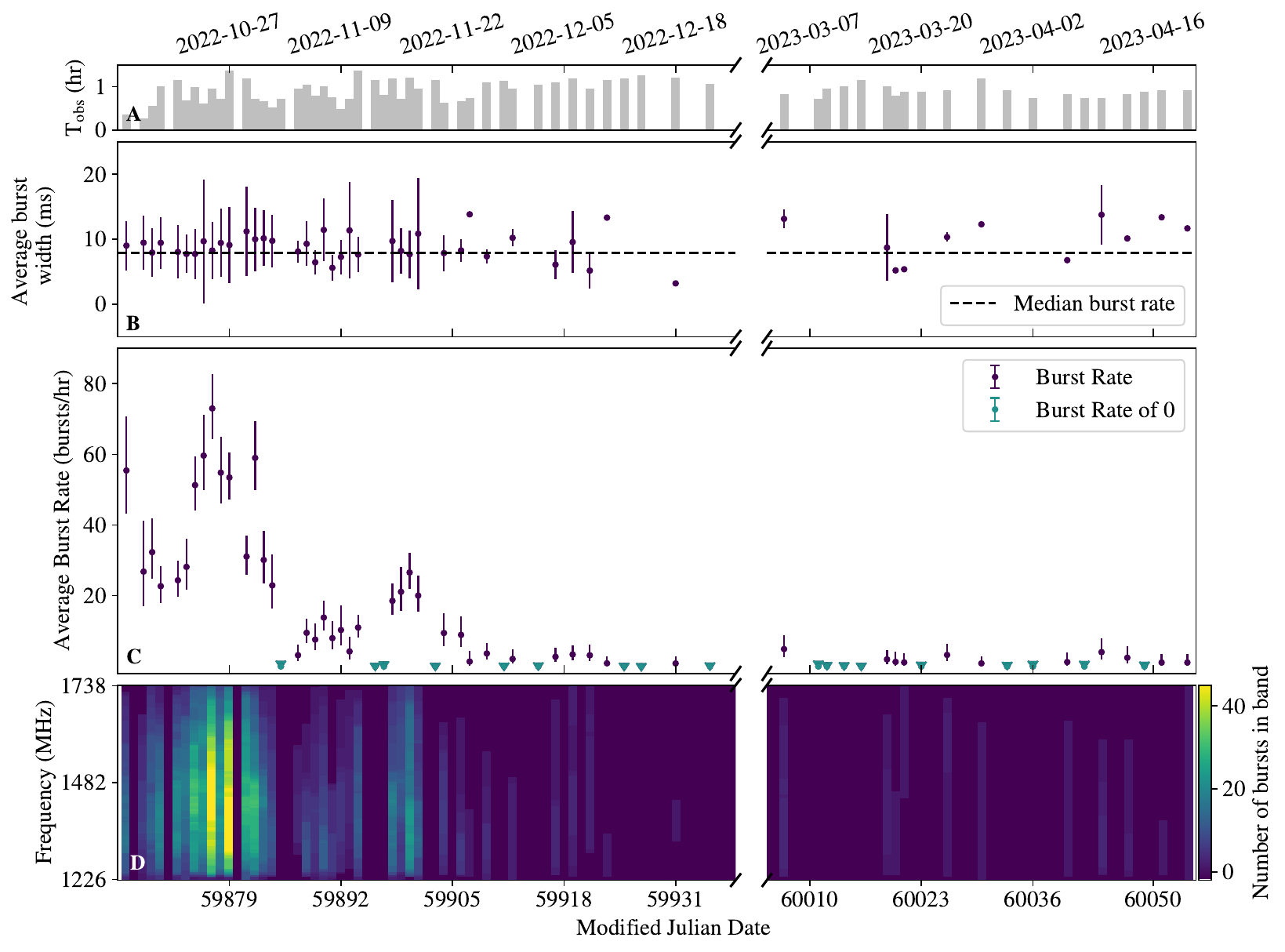}
    \caption{The activity of \FRB during our observing campaign. The error bars on the burst rate represent the Poissonian uncertainty, with a 1-$\sigma$ confidence interval. Panel~A shows the duration of each observation, $T_{\text{obs}}$, in hours. Panel~B shows the average burst width per observation, where the errors indicate the root-mean-square deviation. Panel~D stacks the frequency band of each burst per observation to illustrate the total number of bursts within each frequency channel.}
    \label{fig:burstrates}
\end{figure*}

We mapped the burst activity and spectral evolution of \FRB bursts in Figure~\ref{fig:burstrates}. The main panel shows the average burst rate per hour, with 1-$\sigma$ Poissonian errors, for each observation against the corresponding MJD. Shortly after the beginning of the observation campaign (MJD~59880), the burst rate peaked at $75^{+10}_{-9}$ bursts per hour, as seen by NRT. The burst rate was highly time-variable, with consecutive observations fluctuating by up to 90 per cent. The gap in observations between MJD~59935 and MJD~60007 is due to a holiday break followed by maintenance. In Panel~A, we show the duration of each observation in hours. Despite our observations having more or less the same duration throughout the campaign, we witness a significant drop in the burst rate around MJD~59910. The burst rate rapidly peaked and then gradually diminished over a period of $\sim80$ days, displaying variations in burst rate during this period.

Panel~D of Figure~\ref{fig:burstrates} is a density plot displaying a superposition of the frequency extent of each individual burst for every observation. When multiple bursts overlap, the coloured lines appear stronger. There is no obvious correlation with the burst rate and the location of the bursts in frequency space.

\section{Discussion}\label{sec:Disc}
\subsection{Burst morphology}
Most of the bursts we detected are narrowband, as is often the case for repeating FRBs \citep[e.g.,][]{andersen2023chime}. A few bright bursts in our sample show `microshots', extremely short-time-scale, broader-band structures. These events are rare ($\lesssim1$\% of the bursts in our sample show this), and we have only identified them in high-S/N bursts that also show millisecond-scale sub-bursts as well. The best examples in our sample were previously discussed in detail by \citet{hewitt2023dense}, who also showed that the occurrence of microshots is clustered in two high-fluence bursts (B411 and B491 in our naming convention). Such microshots have previously been observed superimposed on millisecond-duration bursts from repeaters FRB~20180916B \citep{nimmo2021highly} and FRB~20200120E \citep{nimmo2023burst}. In the case of FRB~20121102A, isolated microshots were seen at high frequencies \citep[$3.9-9.3$\,GHz;][]{snelders2023detection}. Overall, microshots appear to be a rare phenomenon in repeating FRBs, compared to the more typical millisecond-duration bursts, and they may have a different emission mechanism that is triggered by the same underlying event that creates the bursts in general \citep{hewitt2023dense}. 

The NRT bandpass is calculated using the NRT noise diode scans, and is visible in the top histogram of Figure~\ref{fig:bandwidth_vs_central_freq}. It shows greater sensitivity at lower frequencies, potentially (partially) explaining the higher occurrence of bursts at lower frequencies. However, these lower frequencies are also more susceptible to RFI, significantly influencing the bandpass. Therefore, the bandpass may also somewhat reflect the higher levels of RFI, instead of the innate higher sensitivity. The apparent preference for burst emission at lower frequencies in the NRT observing band agrees with observations of \FRB by the Five-hundred-meter Aperture Spherical Telescope \citep[FAST;][]{zhang2023fast} and the Allen Telescope Array \citep[ATA;][]{sheikh2024characterization}. The top histogram of Figure~\ref{fig:bandwidth_vs_central_freq} also indicates that bursts at these lower frequencies were generally more energetic. Observations by FAST and ATA were conducted with an observing band that stretches to slightly lower frequencies ($1000-1500$\,MHz and $900-2244$\,MHz for FAST and ATA, respectively) than our NRT data. The central frequency of the bursts was found to be drifting down by 6.12 $\pm$ 0.76\,MHz per day \citep{sheikh2024characterization}. Furthermore, the majority of burst emission seems to occur just below the NRT observing band which, in combination with the decreasing central frequencies of the bursts, gives rise to a more rapid drop in burst rate observed by NRT (around MJD~59900), compared to FAST and ATA.

\subsection{Energetics}
Plotted in Panel~D of Figure~\ref{fig:energydistribution} is the cumulative spectral energy distribution, where the three representative completeness thresholds deviate from the start of the power law by approximately $10^{0.61}, 10^{0.54},$ and $10^{0.47}$ erg/Hz. This could indicate that the spectral energy distribution flattens at lower energies. \citet{zhang2023fast} observe a bimodality of the energy distribution for \FRB, and other repeating FRBs, like FRB~20121102A and FRB~20201124A, also show this bimodal energy distribution \citep[e.g.,][]{aggarwal2021comprehensive, li2021bimodal, jahns2023frb}. Due to our limited sensitivity compared with FAST, we are likely only sampling the tail of the distribution at lower energies, which is why the distribution also seems to flatten at lower energies and which is likely why the completeness thresholds and start of the power-law fit are misaligned. We do not attempt to model a low-energy turnover in the burst rate.

The cumulative spectral energy distribution also flattens at higher energies, showing that higher-energy FRBs occur more often than would be expected based on an extrapolation from a simple power law. This aspect of \FRB's behaviour will be discussed in more detail by \citet{R117paper}, who incorporate additional data from several other telescopes as well as more advanced modelling.

\subsection{Wait-times}
Figure~\ref{fig:waittimes} shows the burst wait-time distribution. Similar to many other repeaters \citep[e.g.,][]{li2021bimodal,hewitt2022arecibo,nimmo2023burst}, the distribution seen from \FRB is bimodal. We find that the longer-timescale (right) peak of the wait-time distribution lies at 67.03\,s, which is at longer timescales than the equivalent wait-time peak seen from FAST observations of \FRB \citep[18.05\,s;][]{zhang2023fast}. The fluctuating burst rate, 20-fold sensitivity difference between the NRT and FAST, and the fact that FAST observations are typically half as long on average, all influence the position of this peak. \citet{aggarwal2021comprehensive}, e.g., have shown how this right peak shifts to shorter timescales as more bursts are detected in a constant-length observation. 

The shorter-timescale (left) peak in the wait-time distribution (33.40\,ms) is comparable not only to what has been found by other telescopes observing this source, e.g. 50.98\,ms \citep{zhang2023fast}, but also to other sources such as FRB~20201124A (51\,ms in \citeauthor{zhang2022fast} \citeyear{zhang2022fast} and 39\,ms in \citeauthor{xu2022fast} \citeyear{xu2022fast}), and FRB~20121102A \citep[24\,ms in ][]{hewitt2022arecibo}. Note that we classify candidates as separate bursts if the intensity between the peaks returns to the noise level. Other authors have used different methods to distinguish between a single multi-component burst and multiple bursts, which could result in different peaks in the wait-time distribution \citep[e.g., see][for a discussion and distinction]{jahns2023frb}. The increased likelihood of detecting a burst very shortly ($<$ 100\,ms) after another burst, appears to be universal among the most active repeaters, with the exception of FRB~20200120E, the globular cluster repeater. FRB~20200120E also shows a potential bimodality in its wait-time distribution, but rather than at tens of milliseconds, the short-wait-time peak is at 0.94\,seconds \citep{nimmo2023burst}. Future studies should aim to constrain whether the short-wait-time peak is constant with time for a particular source, and how the timescale compares between sources. Additionally, the existing datasets need to be reanalyzed with consistent assumptions above what constitutes a sub-burst versus a separate burst. 

The timescales of the short-wait-time peaks seen from repeaters might be associated with the physical size of the region where bursts are generated, or perhaps the time required for perturbations to propagate through this region \citep{nimmo2023burst, totani2023fast}. Such a bimodality in the wait-time distribution has also been seen in short X-ray bursts from magnetars \citep[e.g.,][]{huppenkothen2015dissecting}. While \citet{huppenkothen2015dissecting} suggest that these wait-times align with both repeated crust failure and magnetospheric reconnection models, \citet{wadiasingh2020fast} propose that the shorter timescales observed in the wait-time distributions of magnetars are more likely associated with crustal magnetar oscillations. Another alternative that might explain the presence of the short-wait-time peak is the occurrence of aftershocks following a starquake within a neutron star crust \citep{totani2023fast}. If consecutive bursts were indeed the result of aftershocks, one might expect a trend where fluence of the first burst is consistently higher; however, this relation has not been found in this work, or in previous studies \citep[e.g.][]{gourdji2019sample}.

\subsection{Burst rate}
High variability in the burst rate, as illustrated for \FRB in Figure~\ref{fig:burstrates}, has also been seen in other repeating FRBs, namely: FRB~20121102, FRB~20200120E, FRB~20201124A, and FRB~20240114A \citep[e.g.,][]{li2021bimodal, nimmo2023burst, lanman2022sudden, shin2024chime}. This variability is not only seen in other FRBs but also in magnetars and pulsars that emit giant pulses. Magnetars can exhibit outbursts during which they emit numerous short X-ray bursts within a short time frame, typically ranging from tens to hundreds of bursts per hour \citep{gavriil2004comprehensive, israel2008swift, van2012sgr}. An excellent example of this high variablity was seen in 2020 from the magnetar, SGR~1935+2154, when it generated a millisecond-duration radio burst with a luminosity comparable to the lowest-luminosity FRBs \citep{chime2020bright, bochenek2020fast}. During its outburst phase, the X-ray burst rate was approximately 720 bursts per hour \citep{fletcher2020fermi, palmer2020forest, younes2020nicer}, but rapidly declined to approximately 29 bursts per hour within a mere 3-hour period. In the context of giant pulse emitters, significant variations in the rate of giant pulses have also been observed across different observing epochs. For instance, in the case of the Crab pulsar (PSR~B0531+21), the rate of giant pulses with high fluence ($>$ 130\,Jy\,ms) has been found to vary by up to a factor of five between consecutive observing days \citep{bera2019super}.

Motivated by the lack of clear bimodality in the repetition rate distribution of FRB sources \citep[repeaters versus non-repeaters;][]{andersen2023chime}, and the potential of a correlation between repetition rate and burst width \citep{connor2020beaming}, we searched for a correlation between the burst rate and burst width in our sample. Such a correlation across a population of sources could be attributed to differences in the beaming angle, whereby FRBs with a broader beaming angle would exhibit higher observed repetition rates. The top panel of Figure~\ref{fig:burstrates} shows the median burst width per observation against the burst rate of that observation. We see no correlation between the median burst width and burst rate on a per-observation basis, concluding that for our sample of bursts from \FRB the burst width is independent of activity rate. While this correlation does not exist for our sample of bursts from \FRB, it might still exist across a larger sample of different sources and we encourage future works to investigate this. 

The total wait-time distribution, considered over the entire observing campaign, shows clustering, but individual observations on hour-long timescales appear Poissonian (see Appendix~\ref{sec:waittimessingle}). To investigate further, we applied an identical Weibull analysis to the wait-time distribution of the FAST data presented in \citet{zhang2023fast}. These observations have higher burst rates, and have shorter observation durations ($\sim$30 minutes). Using the total set of observations, we obtained a shape parameter of $k = 0.93 \pm 0.003 \pm 0.01$. Despite FAST's increased sensitivity, higher burst rate, and the shorter duration of the observations, it also demonstrates clustering in the wait-times.

\subsection{DM and Drift}
\begin{figure*}
  \begin{subfigure}{0.48\textwidth}
    \includegraphics[width=\linewidth]{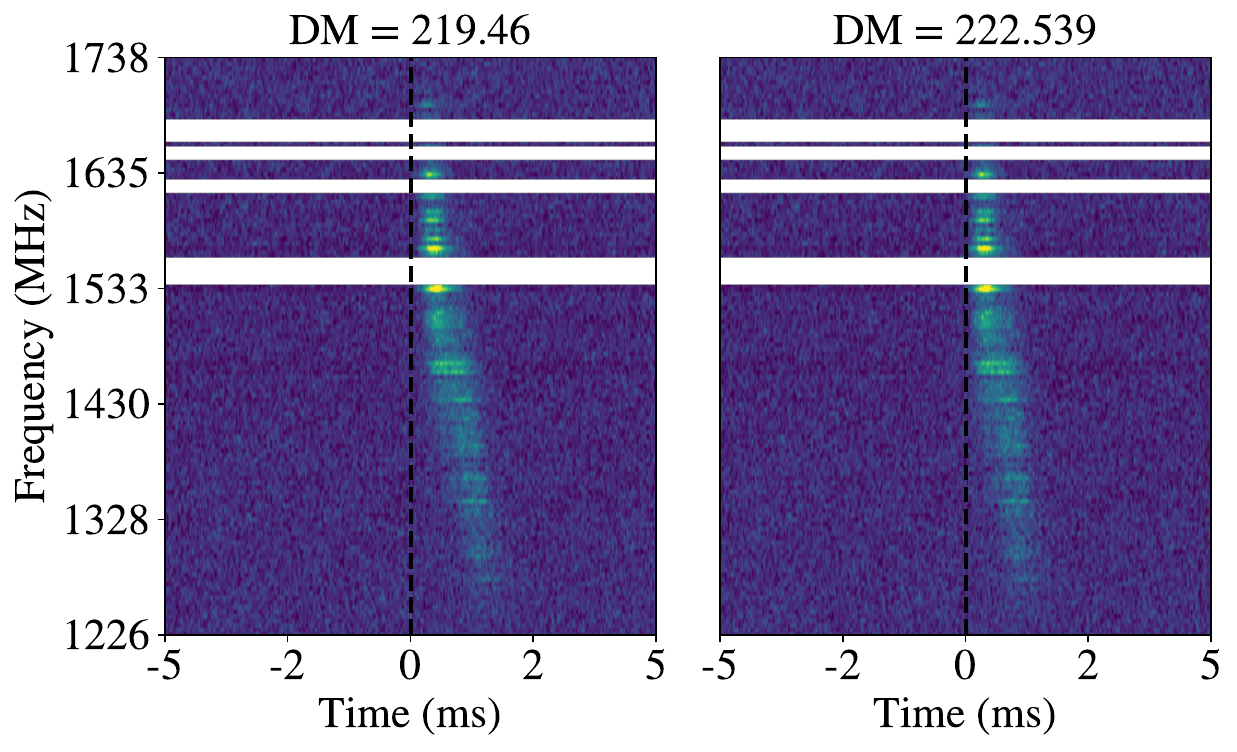}
    \caption{Dynamic spectrum of burst B664 dedispersed to a DM of 219.46\,pc\,cm$^{-3}$ and to a DM of 222.539~pc\,cm$^{-3}$.} \label{fig:comp_a}
  \end{subfigure}%
  \hspace*{\fill}   
  \begin{subfigure}{0.48\textwidth}
    \includegraphics[width=\linewidth]{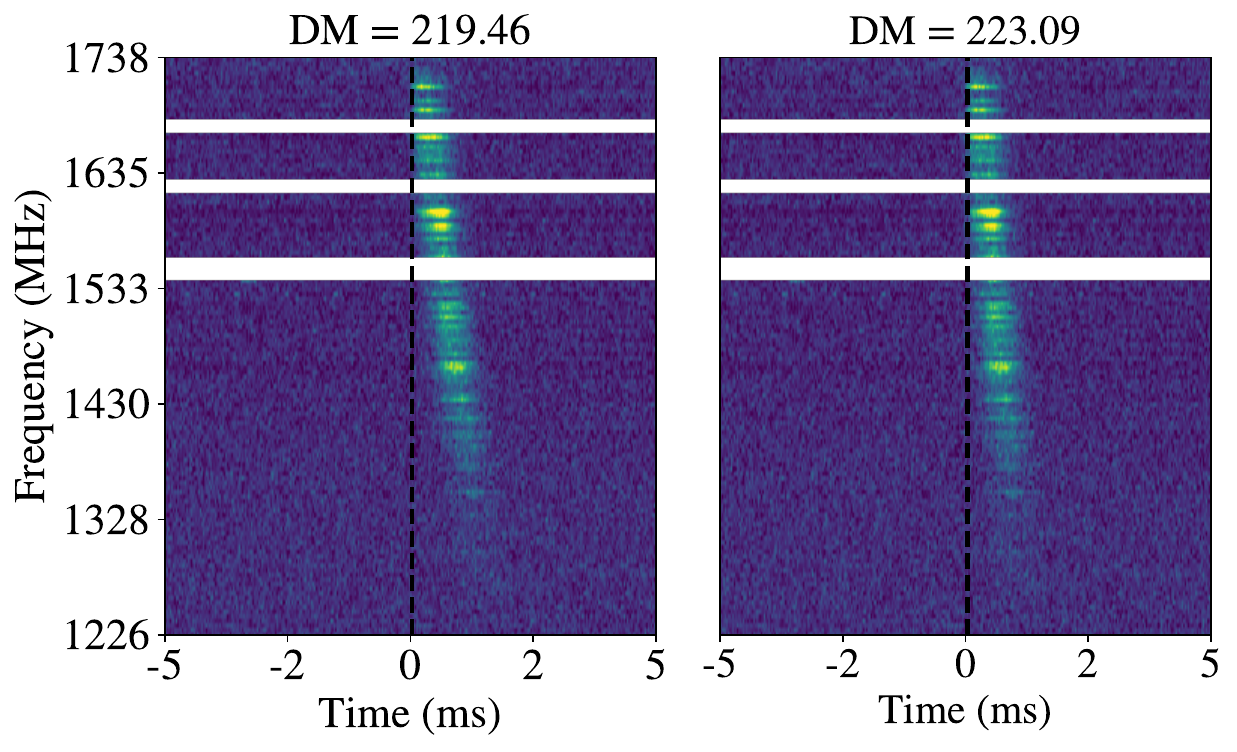}
    \caption{Dynamic spectrum of burst B669 dedispersed to a DM of 219.46\,pc\,cm$^{-3}$ and to a DM of 223.090\,pc\,cm$^{-3}$.} \label{fig:comp_b}
  \end{subfigure}%
\caption{Dynamic spectra of bursts B664 and B669, dedispersed to a DM of 219.46\,pc\,cm$^{-3}$ and to the DM optimized for maximising coherent power in each burst. The black dotted line marks the onset of each burst. Visual inspection suggests consistency with a DM of 219.46\,pc\,cm$^{-3}$ for both bursts.} 
\label{fig:comp_dms}
\end{figure*}

Figure~\ref{fig:DM} shows the DM of bursts with S/N$>120$ and fluctuation frequencies exceeding 2.5~ms$^{-1}$, where we observed an apparent increase in the DM from MJD~59900 to MJD~59920. However, this apparent increase in the DM was not corroborated when we looked at the total drift rate of a set of bright (S/N$>30$) bursts, shown in Figure~\ref{fig:drift_bursts}. We therefore conclude that we have incorporated the sad-trombone drift or intra-burst drift into the DM calculation when determining the DMs of bursts B664 and B669. Figure~\ref{fig:comp_dms} shows both B664 and B669 dedispersed to a DM of 219.46\,pc\,cm$^{-3}$ and to the DM optimized for maximising coherent power in each burst. Here, both bursts appear consistent with a DM of 219.46\,pc\,cm$^{-3}$. Although we do not find any large DM variations over time, \citep{hewitt2023dense} reported robust evidence for DM variations on the level of $\sim0.3$\,pc\,cm$^{-3}$ on the timescale of weeks -- using broad-band microshots to unambiguously determine the DM accurately and precisely.

We find a mean drift rate for \FRB of $-8.8\,$MHz\,ms$^{-1}$, which is a combination of the excess DM beyond 219.46\,pc\,cm$^{-3}$, the drift induced by the sad-trombone effect, and the intra-burst drift. Our mean drift is more than an order-of-magnitude smaller than those quoted for FRB~20121102A in \citet{hessels2019frb} ($\sim-200\,$MHz\,ms$^{-1}$), where the sad trombone effect dominated the drift rate. Our mean drift rate is comparable with the drift rate from FRB~20180916B quoted in \citet{pastor2021chromatic} at similar frequencies ($-39 \pm 7\,$MHz\,ms$^{-1}$). Note that our quoted drift rate is likely smaller, due to the fact that we have fixed our DM to a constant value. Our drift rate is also comparable to FRB~20201124A, where \citet{zhou2022fast} quotes a mean single component drift rate of $-61 \pm 9\,$MHz\,ms$^{-1}$, and a mean multi-component drift rate of $-21 \pm 4\,$MHz\,ms$^{-1}$.

\section{Conclusions \& Future work}\label{sec:Concl}
We performed a study of the repeating \FRB using the NRT, where a total of 61 hours of observations were conducted over the course of 6 months. Employing the new \texttt{CATCH} pipeline, a total of 696\,bursts were detected, with an observed event rate up to $75^{+10}_{-9}$\,bursts per hour (above fluence 0.59\,Jy\,ms), demonstrating the remarkably high activity of the source, which has also been noted by other studies \citep[e.g.,][]{zhang2023fast, feng2022extreme, mckinven2022nine}. \FRB shows various features which appear to be common among (hyper)active repeaters:

\begin{itemize}
    \item The wait-time distribution is bimodal, showing peaks at 33.4\,ms and 67.0\,s. The short-wait-time peak at 33.4\,ms is roughly consistent between repeating FRB sources, to within a factor $\sim2$ \citep[e.g.,][]{hewitt2022arecibo}.
    \item The cumulative spectral energy distribution flattens towards the highest energies ($E_\nu \gtrsim 10^{31}$~erg/Hz), as also seen for another hyperactive repeater \citep{kirsten2024link}. We will analyse and discuss this in more detail in an upcoming paper \citep{R117paper}.
    \item The DM is roughly constant in time when taking into account the confusing factor of `sad-trombone' time-frequency drift ($\delta$DM\,$\lesssim$\,2\,pc\,cm$^{-3}$). Nonetheless, we confirm the results of \citet{hewitt2023dense}, and find that small DM variations, on the order of $\sim0.3$\,pc\,cm$^{-3}$, are occurring.
    \item The bursts show a mean `sad trombone' drift rate of $-8.8\,$MHz\,ms$^{-1}$ (for bursts with S/N$>30$). This is comparable to some known repeaters, but much less compared to others measured in the same frequency range \citep[e.g.,][]{hessels2019frb}.
    \item We observe that more bursts occur towards the bottom of the NRT observing band, at around 1350\,MHz, consistent with quasi-contemporaneous observations from FAST \citep{zhang2023fast} and ATA \citep{sheikh2024characterization}.
    \item While the burst rate within $\sim$1-hr observations appears Poissonian, we find that the rate in the entire observing campaign is better modelled by a Weibull distribution, with shape parameter $k=0.88\,\pm\,0.01\,\pm\,0.01$, showing clustering.
    \item The 16-$\upmu$s time resolution of the data allows us to probe microshots in the bursts, but we find that these occur in only $\lesssim 1$\% of the bursts in our sample. The rare occurrence of microshots has also been seen in other repeaters \citep[e.g.][]{nimmo2023burst}.
\end{itemize}

The breadth of phenomena displayed by repeating FRBs provides multiple dimensions to quantitatively compare their properties and to constrain theoretical models of their nature. 
\'ECLAT is continuing to monitor repeating FRBs at $1218-1740$\,MHz (and sometimes also at $>2$\,GHz) with approximately weekly cadence, providing a valuable complement to CHIME/FRB's daily monitoring of these sources at $400-800$\,MHz.

\section*{Acknowledgements}

The AstroFlash research group at McGill University, University of Amsterdam, ASTRON, and JIVE is supported by: a Canada Excellence Research Chair in Transient Astrophysics (CERC-2022-00009); the European Research Council (ERC) under the European Union’s Horizon 2020 research and innovation programme (`EuroFlash'; Grant agreement No. 101098079); and an NWO-Vici grant (`AstroFlash'; VI.C.192.045). The \nancay Radio Observatory is operated by the Paris Observatory, associated with the French {\it Centre National de la Recherche Scientifique} (CNRS). We acknowledge financial support from the {\it Programme National de Cosmologie et Galaxies} (PNCG) and {\it Programme National Hautes Energies} (PNHE) of INSU, CNRS, France. K.~N. is an MIT Kavli Fellow.

\section*{Data Availability}

The relevant code and data products for this work will be uploaded on Zenodo at the time of publication.



\bibliographystyle{mnras}
\bibliography{references} 

\begin{thebibliography}{}
\makeatletter
\relax
\def\mn@urlcharsother{\let\do\@makeother \do\$\do\&\do\#\do\^\do\_\do\%\do\~}
\def\mn@doi{\begingroup\mn@urlcharsother \@ifnextchar [ {\mn@doi@} {\mn@doi@[]}}
\def\mn@doi@[#1]#2{\def\@tempa{#1}\ifx\@tempa\@empty \href {http://dx.doi.org/#2} {doi:#2}\else \href {http://dx.doi.org/#2} {#1}\fi \endgroup}
\def\mn@eprint#1#2{\mn@eprint@#1:#2::\@nil}
\def\mn@eprint@arXiv#1{\href {http://arxiv.org/abs/#1} {{\tt arXiv:#1}}}
\def\mn@eprint@dblp#1{\href {http://dblp.uni-trier.de/rec/bibtex/#1.xml} {dblp:#1}}
\def\mn@eprint@#1:#2:#3:#4\@nil{\def\@tempa {#1}\def\@tempb {#2}\def\@tempc {#3}\ifx \@tempc \@empty \let \@tempc \@tempb \let \@tempb \@tempa \fi \ifx \@tempb \@empty \def\@tempb {arXiv}\fi \@ifundefined {mn@eprint@\@tempb}{\@tempb:\@tempc}{\expandafter \expandafter \csname mn@eprint@\@tempb\endcsname \expandafter{\@tempc}}}

\bibitem[\protect\citeauthoryear{Agarwal \& Aggarwal}{Agarwal \& Aggarwal}{2020}]{agarwal_aggarwal_2020}
Agarwal D.,  Aggarwal K.,  2020, {devanshkv/fetch: Software release with the manuscript}, \mn@doi{10.5281/zenodo.3905437}, \url {https://doi.org/10.5281/zenodo.3905437}

\bibitem[\protect\citeauthoryear{Agarwal, Aggarwal, Burke-Spolaor, Lorimer  \& Garver-Daniels}{Agarwal et~al.}{2020}]{agarwal2020fetch}
Agarwal D.,  Aggarwal K.,  Burke-Spolaor S.,  Lorimer D.~R.,   Garver-Daniels N.,  2020, Monthly Notices of the Royal Astronomical Society, 497, 1661

\bibitem[\protect\citeauthoryear{Aggarwal, Agarwal, Lewis, Anna-Thomas, Tremblay, Burke-Spolaor, McLaughlin  \& Lorimer}{Aggarwal et~al.}{2021}]{aggarwal2021comprehensive}
Aggarwal K.,  Agarwal D.,  Lewis E.~F.,  Anna-Thomas R.,  Tremblay J.~C.,  Burke-Spolaor S.,  McLaughlin M.~A.,   Lorimer D.~R.,  2021, The Astrophysical Journal, 922, 115

\bibitem[\protect\citeauthoryear{Bera \& Chengalur}{Bera \& Chengalur}{2019}]{bera2019super}
Bera A.,  Chengalur J.~N.,  2019, Monthly Notices of the Royal Astronomical Society: Letters, 490, L12

\bibitem[\protect\citeauthoryear{Bhandari et~al.,}{Bhandari et~al.}{2023}]{bhandari2023constraints}
Bhandari S.,  et~al., 2023, The Astrophysical Journal Letters, 958, L19

\bibitem[\protect\citeauthoryear{Bhusare et~al.,}{Bhusare et~al.}{2022}]{bhusare2022ugmrt}
Bhusare Y.,  et~al., 2022, The Astronomer's Telegram, 15806, 1

\bibitem[\protect\citeauthoryear{Bochenek, Ravi, Belov, Hallinan, Kocz, Kulkarni  \& McKenna}{Bochenek et~al.}{2020}]{bochenek2020fast}
Bochenek C.~D.,  Ravi V.,  Belov K.~V.,  Hallinan G.,  Kocz J.,  Kulkarni S.~R.,   McKenna D.~L.,  2020, Nature, 587, 59

\bibitem[\protect\citeauthoryear{{CHIME/FRB Collaboration}}{{CHIME/FRB Collaboration}}{2020}]{chime2020bright}
{CHIME/FRB Collaboration} 2020, Nature, 587, 54

\bibitem[\protect\citeauthoryear{{CHIME/FRB Collaboration} Andersen et~al.,}{{CHIME/FRB Collaboration} et~al.}{2023}]{andersen2023chime}
{CHIME/FRB Collaboration} Andersen B.~C.,  et~al., 2023, The Astrophysical Journal, 947, 83

\bibitem[\protect\citeauthoryear{Camilo, Reynolds, Johnston, Halpern  \& Ransom}{Camilo et~al.}{2008}]{camilo2008magnetar}
Camilo F.,  Reynolds J.,  Johnston S.,  Halpern J.,   Ransom S.,  2008, The Astrophysical Journal, 679, 681

\bibitem[\protect\citeauthoryear{Chamma, Rajabi, Wyenberg, Mathews  \& Houde}{Chamma et~al.}{2021}]{chamma2021evidence}
Chamma M.~A.,  Rajabi F.,  Wyenberg C.~M.,  Mathews A.,   Houde M.,  2021, Monthly Notices of the Royal Astronomical Society, 507, 246

\bibitem[\protect\citeauthoryear{Clauset, Shalizi  \& Newman}{Clauset et~al.}{2009}]{clauset2009power}
Clauset A.,  Shalizi C.~R.,   Newman M.~E.,  2009, SIAM review, 51, 661

\bibitem[\protect\citeauthoryear{Connor, Miller  \& Gardenier}{Connor et~al.}{2020}]{connor2020beaming}
Connor L.,  Miller M.,   Gardenier D.,  2020, Monthly Notices of the Royal Astronomical Society, 497, 3076

\bibitem[\protect\citeauthoryear{Cordes \& McLaughlin}{Cordes \& McLaughlin}{2003}]{cordes2003searches}
Cordes J.,  McLaughlin M.~A.,  2003, The Astrophysical Journal, 596, 1142

\bibitem[\protect\citeauthoryear{Dai et~al.,}{Dai et~al.}{2019}]{dai2019wideband}
Dai S.,  et~al., 2019, The Astrophysical Journal Letters, 874, L14

\bibitem[\protect\citeauthoryear{Desvignes, Barott, Cognard, Lespagnol, Theureau  et~al.}{Desvignes et~al.}{2011}]{desvignes2011new}
Desvignes G.,  Barott W.~C.,  Cognard I.,  Lespagnol P.,  Theureau G.,   et~al., 2011, in AIP Conference Proceedings. pp 349--350

\bibitem[\protect\citeauthoryear{Feng et~al.,}{Feng et~al.}{2022}]{feng2022extreme}
Feng Y.,  et~al., 2022, The Astronomer's Telegram, 15723, 1

\bibitem[\protect\citeauthoryear{{Fletcher \& Fermi GBM Team}}{{Fletcher \& Fermi GBM Team}}{2020}]{fletcher2020fermi}
{Fletcher \& Fermi GBM Team} 2020, GRB Coordinates Network, 27659, 1

\bibitem[\protect\citeauthoryear{Gajjar et~al.,}{Gajjar et~al.}{2018}]{gajjar2018highest}
Gajjar V.,  et~al., 2018, The Astrophysical Journal, 863, 2

\bibitem[\protect\citeauthoryear{Gavriil, Kaspi  \& Woods}{Gavriil et~al.}{2004}]{gavriil2004comprehensive}
Gavriil F.~P.,  Kaspi V.~M.,   Woods P.~M.,  2004, The Astrophysical Journal, 607, 959

\bibitem[\protect\citeauthoryear{Gopinath et~al.,}{Gopinath et~al.}{2024}]{gopinath2024propagation}
Gopinath A.,  et~al., 2024, Monthly Notices of the Royal Astronomical Society, 527, 9872

\bibitem[\protect\citeauthoryear{Gourdji, Michilli, Spitler, Hessels, Seymour, Cordes  \& Chatterjee}{Gourdji et~al.}{2019}]{gourdji2019sample}
Gourdji K.,  Michilli D.,  Spitler L.,  Hessels J.,  Seymour A.,  Cordes J.,   Chatterjee S.,  2019, The Astrophysical Journal Letters, 877, L19

\bibitem[\protect\citeauthoryear{Gourdji, Rowlinson, Wijers  \& Goldstein}{Gourdji et~al.}{2020}]{gourdji2020constraining}
Gourdji K.,  Rowlinson A.,  Wijers R.~A.,   Goldstein A.,  2020, Monthly Notices of the Royal Astronomical Society, 497, 3131

\bibitem[\protect\citeauthoryear{Hessels et~al.,}{Hessels et~al.}{2019}]{hessels2019frb}
Hessels J.,  et~al., 2019, The Astrophysical Journal Letters, 876, L23

\bibitem[\protect\citeauthoryear{Hewitt et~al.,}{Hewitt et~al.}{2022}]{hewitt2022arecibo}
Hewitt D.,  et~al., 2022, Monthly Notices of the Royal Astronomical Society, 515, 3577

\bibitem[\protect\citeauthoryear{Hewitt et~al.,}{Hewitt et~al.}{2023}]{hewitt2023dense}
Hewitt D.~M.,  et~al., 2023, Monthly Notices of the Royal Astronomical Society, 526, 2039

\bibitem[\protect\citeauthoryear{Hewitt et~al.,}{Hewitt et~al.}{2024}]{hewitt2023milliarcsecond}
Hewitt D.~M.,  et~al., 2024, Monthly Notices of the Royal Astronomical Society, 529, 1814

\bibitem[\protect\citeauthoryear{Huppenkothen et~al.,}{Huppenkothen et~al.}{2015}]{huppenkothen2015dissecting}
Huppenkothen D.,  et~al., 2015, The Astrophysical Journal, 810, 66

\bibitem[\protect\citeauthoryear{Israel et~al.,}{Israel et~al.}{2008}]{israel2008swift}
Israel G.,  et~al., 2008, The Astrophysical Journal, 685, 1114

\bibitem[\protect\citeauthoryear{Jahns et~al.,}{Jahns et~al.}{2023}]{jahns2023frb}
Jahns J.,  et~al., 2023, Monthly Notices of the Royal Astronomical Society, 519, 666

\bibitem[\protect\citeauthoryear{Kirsten, Snelders, Jenkins, Nimmo, Van~den Eijnden, Hessels, Gawro{\'n}ski  \& Yang}{Kirsten et~al.}{2021}]{kirsten2021detection}
Kirsten F.,  Snelders M.,  Jenkins M.,  Nimmo K.,  Van~den Eijnden J.,  Hessels J.,  Gawro{\'n}ski M.,   Yang J.,  2021, Nature Astronomy, 5, 414

\bibitem[\protect\citeauthoryear{Kirsten et~al.,}{Kirsten et~al.}{2022a}]{kirsten2022repeating}
Kirsten F.,  et~al., 2022a, Nature, 602, 585

\bibitem[\protect\citeauthoryear{Kirsten et~al.,}{Kirsten et~al.}{2022b}]{kirsten2022precise}
Kirsten F.,  et~al., 2022b, The Astronomer's Telegram, 15727, 1

\bibitem[\protect\citeauthoryear{Kirsten et~al.,}{Kirsten et~al.}{2024}]{kirsten2024link}
Kirsten F.,  et~al., 2024, Nature Astronomy, 8, 337

\bibitem[\protect\citeauthoryear{Konijn}{Konijn}{2023}]{DavidThesis}
Konijn D.,  2023, Uva/VU Master Thesis

\bibitem[\protect\citeauthoryear{Kumar, Qu  \& Zhang}{Kumar et~al.}{2024}]{kumar2024origins}
Kumar P.,  Qu Y.,   Zhang B.,  2024, arXiv preprint arXiv:2406.01266

\bibitem[\protect\citeauthoryear{Lanman et~al.,}{Lanman et~al.}{2022}]{lanman2022sudden}
Lanman A.~E.,  et~al., 2022, The Astrophysical Journal, 927, 59

\bibitem[\protect\citeauthoryear{{Li}, Li, Zhang, Geng, Song, Huang  \& Yang}{{Li} et~al.}{2019}]{li2019statistical}
{Li} Li L.-B.,  Zhang Z.-B.,  Geng J.-J.,  Song L.-M.,  Huang Y.-F.,   Yang Y.-P.,  2019, arXiv preprint arXiv:1901.03484

\bibitem[\protect\citeauthoryear{Li et~al.,}{Li et~al.}{2021}]{li2021bimodal}
Li D.,  et~al., 2021, Nature, 598, 267

\bibitem[\protect\citeauthoryear{Lorimer \& Kramer}{Lorimer \& Kramer}{2005}]{lorimer2005handbook}
Lorimer D.~R.,  Kramer M.,  2005, Handbook of pulsar astronomy

\bibitem[\protect\citeauthoryear{Lorimer, Matthew, McLaughlin, Narkevic  \& Crawford}{Lorimer et~al.}{2007}]{lorimer2007bright}
Lorimer Matthew McLaughlin M.~A.,  Narkevic D.~J.,   Crawford F.,  2007, Science, 318, 777

\bibitem[\protect\citeauthoryear{Macquart \& Ekers}{Macquart \& Ekers}{2018}]{macquart2018frb}
Macquart J.-P.,  Ekers R.,  2018, Monthly Notices of the Royal Astronomical Society, 480, 4211

\bibitem[\protect\citeauthoryear{Marcote et~al.,}{Marcote et~al.}{2020}]{marcote2020repeating}
Marcote B.,  et~al., 2020, Nature, 577, 190

\bibitem[\protect\citeauthoryear{Margalit, Berger  \& Metzger}{Margalit et~al.}{2019}]{margalit2019fast}
Margalit B.,  Berger E.,   Metzger B.~D.,  2019, The Astrophysical Journal, 886, 110

\bibitem[\protect\citeauthoryear{Margalit, Metzger  \& Sironi}{Margalit et~al.}{2020a}]{margalit2020constraints}
Margalit B.,  Metzger B.~D.,   Sironi L.,  2020a, Monthly Notices of the Royal Astronomical Society, 494, 4627

\bibitem[\protect\citeauthoryear{Margalit, Beniamini, Sridhar  \& Metzger}{Margalit et~al.}{2020b}]{margalit2020implications}
Margalit B.,  Beniamini P.,  Sridhar N.,   Metzger B.~D.,  2020b, The Astrophysical Journal Letters, 899, L27

\bibitem[\protect\citeauthoryear{{McKinven \& CHIME/FRB Collaboration} et~al.}{{McKinven \& CHIME/FRB Collaboration} et~al.}{2022}]{mckinven2022nine}
{McKinven \& CHIME/FRB Collaboration} R.,  et~al., 2022, The Astronomer's Telegram, 15679, 1

\bibitem[\protect\citeauthoryear{Nimmo et~al.,}{Nimmo et~al.}{2021}]{nimmo2021highly}
Nimmo K.,  et~al., 2021, Nature Astronomy, 5, 594

\bibitem[\protect\citeauthoryear{Nimmo et~al.,}{Nimmo et~al.}{2022}]{nimmo2022phasespace}
Nimmo K.,  et~al., 2022, Nature Astronomy, 6, 393

\bibitem[\protect\citeauthoryear{Nimmo et~al.,}{Nimmo et~al.}{2023}]{nimmo2023burst}
Nimmo K.,  et~al., 2023, Monthly Notices of the Royal Astronomical Society, 520, 2281

\bibitem[\protect\citeauthoryear{Niu et~al.,}{Niu et~al.}{2022}]{niu2022repeating}
Niu C.-H.,  et~al., 2022, Nature, 606, 873

\bibitem[\protect\citeauthoryear{Oppermann, Yu  \& Pen}{Oppermann et~al.}{2018}]{oppermann2018non}
Oppermann N.,  Yu H.-R.,   Pen U.-L.,  2018, Monthly Notices of the Royal Astronomical Society, 475, 5109

\bibitem[\protect\citeauthoryear{Ould-Boukattine et~al.,}{Ould-Boukattine et~al.}{2022}]{ould2022bright}
Ould-Boukattine O.,  et~al., 2022, The Astronomer's Telegram, 15817, 1

\bibitem[\protect\citeauthoryear{{Ould-Boukattine}, {Kirsten}  \& s.}{{Ould-Boukattine} et~al.}{in prep.}]{R117paper}
{Ould-Boukattine} O.,  {Kirsten} F.,   s. in prep., A probe of the maximum energy of fast radio bursts through a prolific repeating source

\bibitem[\protect\citeauthoryear{Palmer}{Palmer}{2020}]{palmer2020forest}
Palmer D.~M.,  2020, The Astronomer's Telegram, 13675, 1

\bibitem[\protect\citeauthoryear{Papoulis \& Unnikrishna~Pillai}{Papoulis \& Unnikrishna~Pillai}{2002}]{papoulis2002probability}
Papoulis A.,  Unnikrishna~Pillai S.,  2002, Probability, random variables and stochastic processes.
Boston: McGraw-Hill

\bibitem[\protect\citeauthoryear{Pastor-Marazuela et~al.,}{Pastor-Marazuela et~al.}{2021}]{pastor2021chromatic}
Pastor-Marazuela I.,  et~al., 2021, Nature, 596, 505

\bibitem[\protect\citeauthoryear{Pelliciari et~al.,}{Pelliciari et~al.}{2024}]{pelliciari2024northern}
Pelliciari D.,  et~al., 2024, arXiv preprint arXiv:2405.04802

\bibitem[\protect\citeauthoryear{Perera et~al.,}{Perera et~al.}{2022}]{perera2022detection}
Perera B.,  et~al., 2022, The Astronomer's Telegram, 15734, 1

\bibitem[\protect\citeauthoryear{Petroff, Hessels  \& Lorimer}{Petroff et~al.}{2022}]{petroff2022fast}
Petroff E.,  Hessels J.,   Lorimer D.,  2022, The Astronomy and Astrophysics Review, 30, 1

\bibitem[\protect\citeauthoryear{Pleunis et~al.,}{Pleunis et~al.}{2021a}]{pleunis2021lofar}
Pleunis Z.,  et~al., 2021a, The Astrophysical Journal Letters, 911, L3

\bibitem[\protect\citeauthoryear{Pleunis et~al.,}{Pleunis et~al.}{2021b}]{pleunis2021fast}
Pleunis Z.,  et~al., 2021b, The Astrophysical Journal, 923, 1

\bibitem[\protect\citeauthoryear{Rajabi, Chamma, Wyenberg, Mathews  \& Houde}{Rajabi et~al.}{2020}]{rajabi2020simple}
Rajabi F.,  Chamma M.~A.,  Wyenberg C.~M.,  Mathews A.,   Houde M.,  2020, Monthly Notices of the Royal Astronomical Society, 498, 4936

\bibitem[\protect\citeauthoryear{Rajwade et~al.,}{Rajwade et~al.}{2022}]{rajwade2022detection}
Rajwade K.,  et~al., 2022, The Astronomer's Telegram, 15791, 1

\bibitem[\protect\citeauthoryear{Ransom}{Ransom}{2011}]{ransom2011presto}
Ransom S.,  2011, Astrophysics source code library, pp ascl--1107

\bibitem[\protect\citeauthoryear{Ravi et~al.,}{Ravi et~al.}{2022}]{ravi2022host}
Ravi V.,  et~al., 2022, Monthly Notices of the Royal Astronomical Society, 513, 982

\bibitem[\protect\citeauthoryear{Ravi et~al.,}{Ravi et~al.}{2023}]{ravi2023deep}
Ravi V.,  et~al., 2023, The Astrophysical Journal Letters, 949, L3

\bibitem[\protect\citeauthoryear{Seymour, Michilli  \& Pleunis}{Seymour et~al.}{2019}]{seymour2019dm_phase}
Seymour A.,  Michilli D.,   Pleunis Z.,  2019, Astrophysics Source Code Library, pp ascl--1910

\bibitem[\protect\citeauthoryear{Sheikh et~al.,}{Sheikh et~al.}{2022}]{sheikh2022bright}
Sheikh S.,  et~al., 2022, The Astronomer's Telegram, 15735, 1

\bibitem[\protect\citeauthoryear{Sheikh et~al.,}{Sheikh et~al.}{2024}]{sheikh2024characterization}
Sheikh S.~Z.,  et~al., 2024, Monthly Notices of the Royal Astronomical Society, 527, 10425

\bibitem[\protect\citeauthoryear{Shin, Collaboration  et~al.}{Shin et~al.}{2024}]{shin2024chime}
Shin K.,  Collaboration C.,   et~al., 2024, The Astronomer's Telegram, 16420, 1

\bibitem[\protect\citeauthoryear{Snelders et~al.,}{Snelders et~al.}{2023}]{snelders2023detection}
Snelders M.,  et~al., 2023, Nature Astronomy, 7, 1486

\bibitem[\protect\citeauthoryear{Spitler et~al.,}{Spitler et~al.}{2016}]{spitler2016repeating}
Spitler L.,  et~al., 2016, Nature, 531, 202

\bibitem[\protect\citeauthoryear{Thornton et~al.,}{Thornton et~al.}{2013}]{thornton2013population}
Thornton D. e.~a.,  et~al., 2013, Science, 341, 53

\bibitem[\protect\citeauthoryear{Totani \& Tsuzuki}{Totani \& Tsuzuki}{2023}]{totani2023fast}
Totani T.,  Tsuzuki Y.,  2023, Monthly Notices of the Royal Astronomical Society, 526, 2795

\bibitem[\protect\citeauthoryear{Van~der Horst et~al.,}{Van~der Horst et~al.}{2012}]{van2012sgr}
Van~der Horst A.,  et~al., 2012, The Astrophysical Journal, 749, 122

\bibitem[\protect\citeauthoryear{Wadiasingh \& Chirenti}{Wadiasingh \& Chirenti}{2020}]{wadiasingh2020fast}
Wadiasingh Z.,  Chirenti C.,  2020, The Astrophysical Journal Letters, 903, L38

\bibitem[\protect\citeauthoryear{Wang, Luo, Yue, Chen, Lee  \& Xu}{Wang et~al.}{2018}]{wang2018frb}
Wang W.,  Luo R.,  Yue H.,  Chen X.,  Lee K.,   Xu R.,  2018, The Astrophysical Journal, 852, 140

\bibitem[\protect\citeauthoryear{Xu et~al.,}{Xu et~al.}{2022}]{xu2022fast}
Xu H.,  et~al., 2022, Nature, 609, 685

\bibitem[\protect\citeauthoryear{Younes et~al.,}{Younes et~al.}{2020}]{younes2020nicer}
Younes G.,  et~al., 2020, The Astrophysical Journal Letters, 904, L21

\bibitem[\protect\citeauthoryear{Zhang et~al.,}{Zhang et~al.}{2022}]{zhang2022fast}
Zhang Y.-K.,  et~al., 2022, Research in Astronomy and Astrophysics, 22, 124002

\bibitem[\protect\citeauthoryear{Zhang et~al.,}{Zhang et~al.}{2023}]{zhang2023fast}
Zhang Y.-K.,  et~al., 2023, The Astrophysical Journal, 955, 142

\bibitem[\protect\citeauthoryear{Zhou et~al.,}{Zhou et~al.}{2022}]{zhou2022fast}
Zhou D.,  et~al., 2022, Research in Astronomy and Astrophysics, 22, 124001

\makeatother
\end{thebibliography}




\clearpage
\appendix

\section{The Classification Algorithm and Transient Candidate Handler} \label{sec:catch}
Ideally, each FRB candidate from the search pipeline is validated by eye. However, conducting a manual review of tens of thousands of candidates, generated from tens of hours of observations, is not feasible or reproducible. In most FRB searches, convolutional neural network classifiers (e.g., \texttt{FETCH}) have been employed to greatly reduce the number of candidates requiring manual inspection. However, these classifiers are not perfect (especially when not retrained) and occasionally misclassify real FRBs. 

In an attempt to address this issue, we developed a classifier algorithm called "The Classification Algorithm and Transient Candidate Handler" (\texttt{CATCH}, for short). \texttt{CATCH} parses all candidates created by \texttt{Heimdall} that fulfill \textit{any} of the following criteria: i)  \texttt{FETCH}'s deep-learning models `A' to `H' indicate a probability larger than 0.5 that the candidate is astrophysical; ii) candidates that have a subband S/N larger than 200; iii) candidates identified based on their brightness distribution in the DM-time space through a method we will refer to as the `bow tie' method \citep{DavidThesis}. 

\subsection{Identification of the bow tie feature}
\begin{figure}
    \includegraphics[width=0.9\columnwidth]{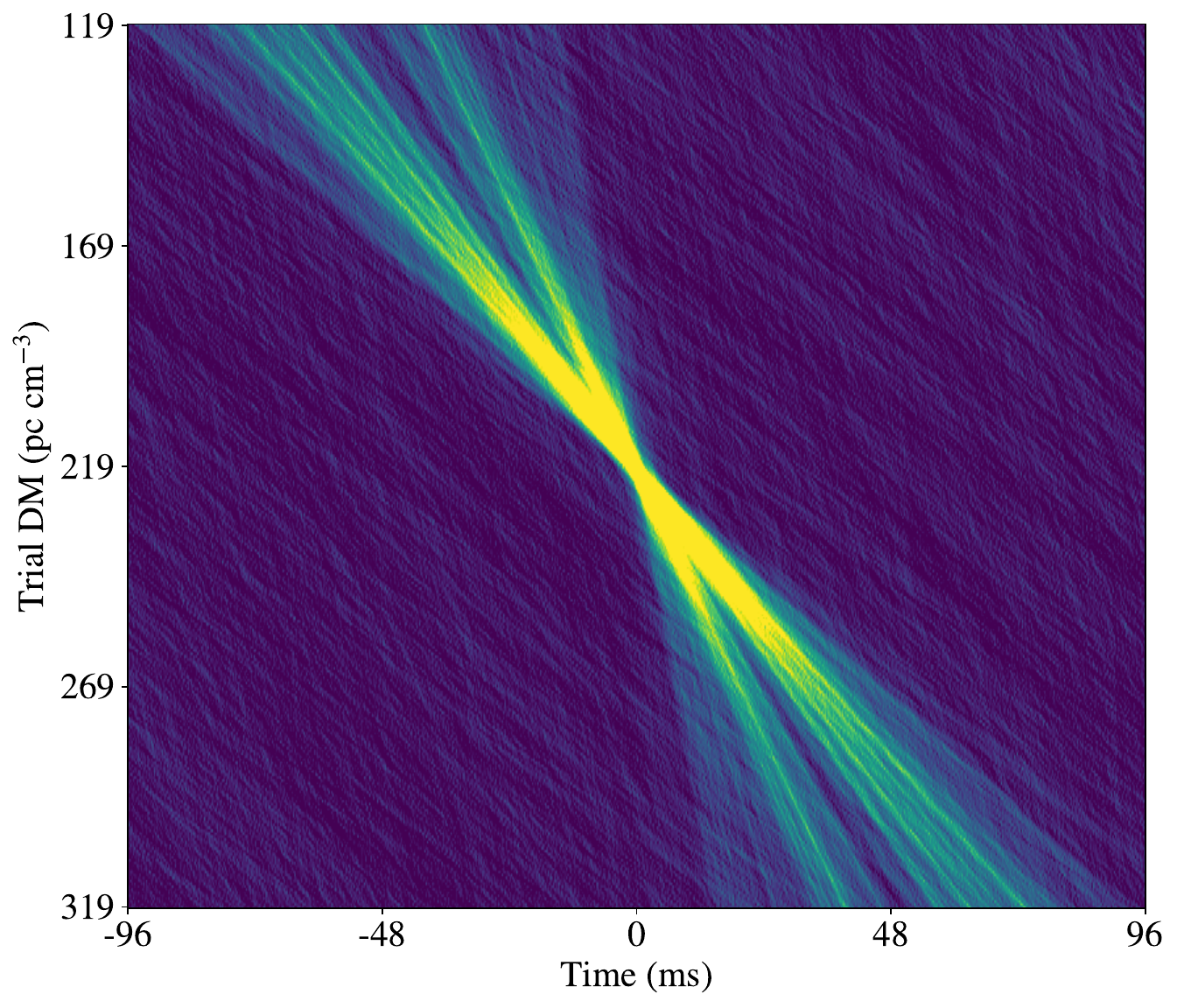}
    \caption{Frequency-integrated burst brightness as a function of DM and time for burst B049. The defining `bow tie'-shaped feature of an impulsive and dispersive astrophysical signal is clearly visible. Fine `spokes' are also visible in the bow tie, indicating time-frequency structure in the dynamic spectrum.}
    \label{fig:FRB_dm_time}
\end{figure}

The bow tie method tries to classify candidates based on their brightness distribution in the DM-time spectrum, i.e. the frequency-integrated burst brightness as a function of DM and time. The DM-time spectrum is created by dedispersing the dynamic spectrum of a burst for a range of trial DM values, and then integrating across all frequencies. In this parameter space, an astrophysical transient with a large non-zero DM, gives rise to a symmetric `bow tie'-like shape, as visible in Figure~\ref{fig:FRB_dm_time}. 

Terrestrial RFI is not significantly dispersed (though it can be frequency-swept), effectively allowing for the identification of astrophysical candidates based on the presence of the bow tie shape in the DM-time spectrum. In an ideal case, the bow tie can be identified by comparing the intensity at the centre of the bow tie (at the true DM) with the intensity at the edges (at some other trial DM). However, this comparison is difficult because the bow tie can vary in size, intensity, opening angle, and pointing angle. To ensure consistency in our comparison, given the different spectro-temporal features of bursts, we first vertically align the bow tie, after which we estimate the size of the bow tie based on the duration and bandwidth of the burst in the dynamic spectrum (frequency-time space).

\begin{figure}
    \centering
    \includegraphics[width=1\columnwidth]{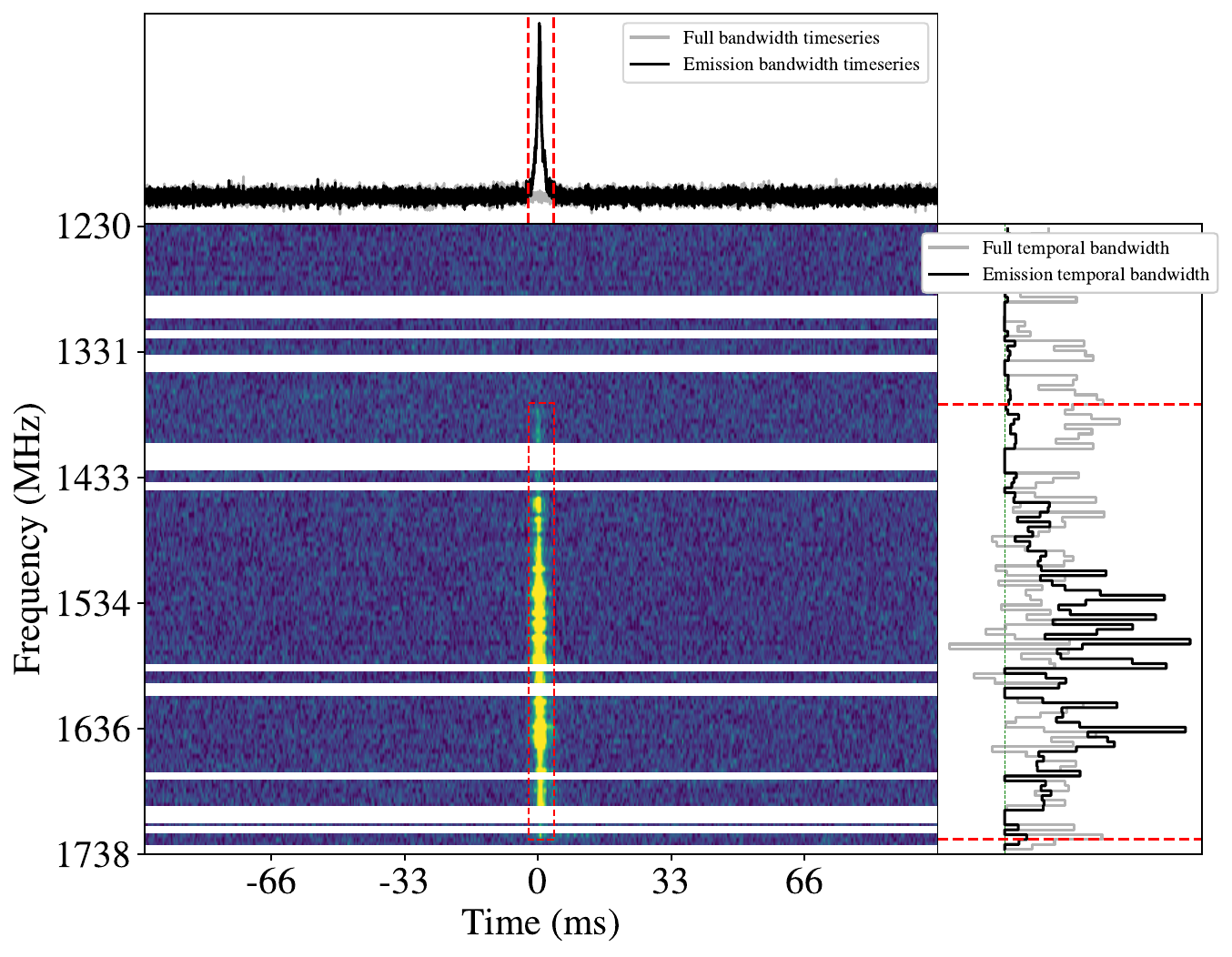}
    \caption{Dynamic spectrum of burst B049. A box has been placed surrounding the highest intensity areas, and is plotted as the red dashed rectangle. The horizontal white bands are the zapped frequency channels, which complicates the process of determining the frequency range of the box. Added to the top and the side of the dynamic spectrum are the one dimensional integrated spectra, where the dashed red lines again indicate the location of the box.}
    \label{fig:B206}
\end{figure}

The bow tie can be vertically aligned by correcting for the temporal delay between the highest and lowest frequency of the burst caused by the dispersion. These frequencies can be obtained by placing a box around the burst in the dynamic spectrum, after masking those channels contaminated by RFI. The optimal location and size of this box are determined by maximising the summed intensity divided by the square root of its area. An example of an optimised box is shown in \ref{fig:B206}. 

\subsection{Bow tie size}
The next step involves determining the size of the bow tie, primarily influenced by the frequency extent and temporal width of the burst. Bursts with large frequency ranges or narrow temporal widths have their frequency-integrated intensity significantly affected by slight variations in DM, resulting in a smaller bow tie. We can calculate this change in intensity due to different DM values per burst and use it to characterise the size of the bow tie. 

Due to dispersion, there is a frequency-dependent time-smearing effect \citep{cordes2003searches}. This is equal to 

\begin{equation}
    \Delta t_{\rm{DM}} = 8.3 \cdot \upmu \text{s} \cdot \text{DM} \cdot \Delta \nu_{\text{MHz}} \cdot \nu_{\text{GHz}}^{-3},
\end{equation}

\noindent for a frequency extent $\Delta \nu_{\text{MHz}}$, an observing frequency $\nu_{\text{GHz}}$, and a dispersion measure DM. Using a DM value that is incorrect by an amount $\delta$DM increases the temporal width of the pulse by 

\begin{equation}
    \Delta t_{\delta \text{DM}} = \Delta t_{\text{DM}} (\delta \text{DM}/\text{DM}).
\end{equation}

By combining these equations, an expression for the time smearing as a function of incorrect DM value can be found:

\begin{equation}
    \Delta t_{\delta \text{DM}} = 8.3 \ \upmu \text{s} \cdot \Delta \nu_{\text{MHz}} \cdot \nu_{\text{GHz}}^{-3} \cdot \delta \text{DM}.
\end{equation}

Reordering terms results in an equation for $\delta$DM as a function of the temporal smearing: 
\begin{equation}
     \delta \text{DM} = \Delta t_{\delta \text{DM}} / ( 8.3 \ \upmu \text{s} \cdot \Delta \nu_{\text{MHz}} \cdot \nu_{\text{GHz}}^{-3}).
\end{equation}

Since the total power of the burst remains constant, time smearing that will increase the width of the burst by a factor N will decrease the  per-pixel intensity in the DM-time spectrum by a factor sqrt(N). By assuming $\Delta t_{\delta \text{DM}} = 10\times t_{\text{burst}}$, we can compute the error associated with the DM required to achieve this temporal width. Thus, for each burst, we can calculate the number of DM trials where the bow tie width will be smeared by a factor of ten and the intensity reduced by a factor of $\sqrt{10}$. We define the size of the symmetrical bow tie as the number of DM units on both sides that reduce the intensity with a factor of $\sqrt{10}$.

To identify the bow tie, we compare the intensity in the centre of the bow tie with the intensity at the top and bottom of the bow tie. The width of the box is equal to the temporal width of the burst in the dynamic spectrum, and the height of the box is calculated using the above equations where $\Delta t_{\delta \text{DM}} = 3\times t_{\text{burst}}$, to take into account the smearing of the burst. The amount of DM units between the centre of the box and the top and bottom has been calculated using $10\times t_{\text{burst}}$. Here, the box is slid across the spectrum at the calculated trial DM using a sliding window of $10\times t_{\text{burst}}$ ms, where the box with the highest intensity is selected for comparison with the box in the centre. 

\begin{figure*}
  \begin{subfigure}{0.48\textwidth}
    \includegraphics[width=\linewidth]{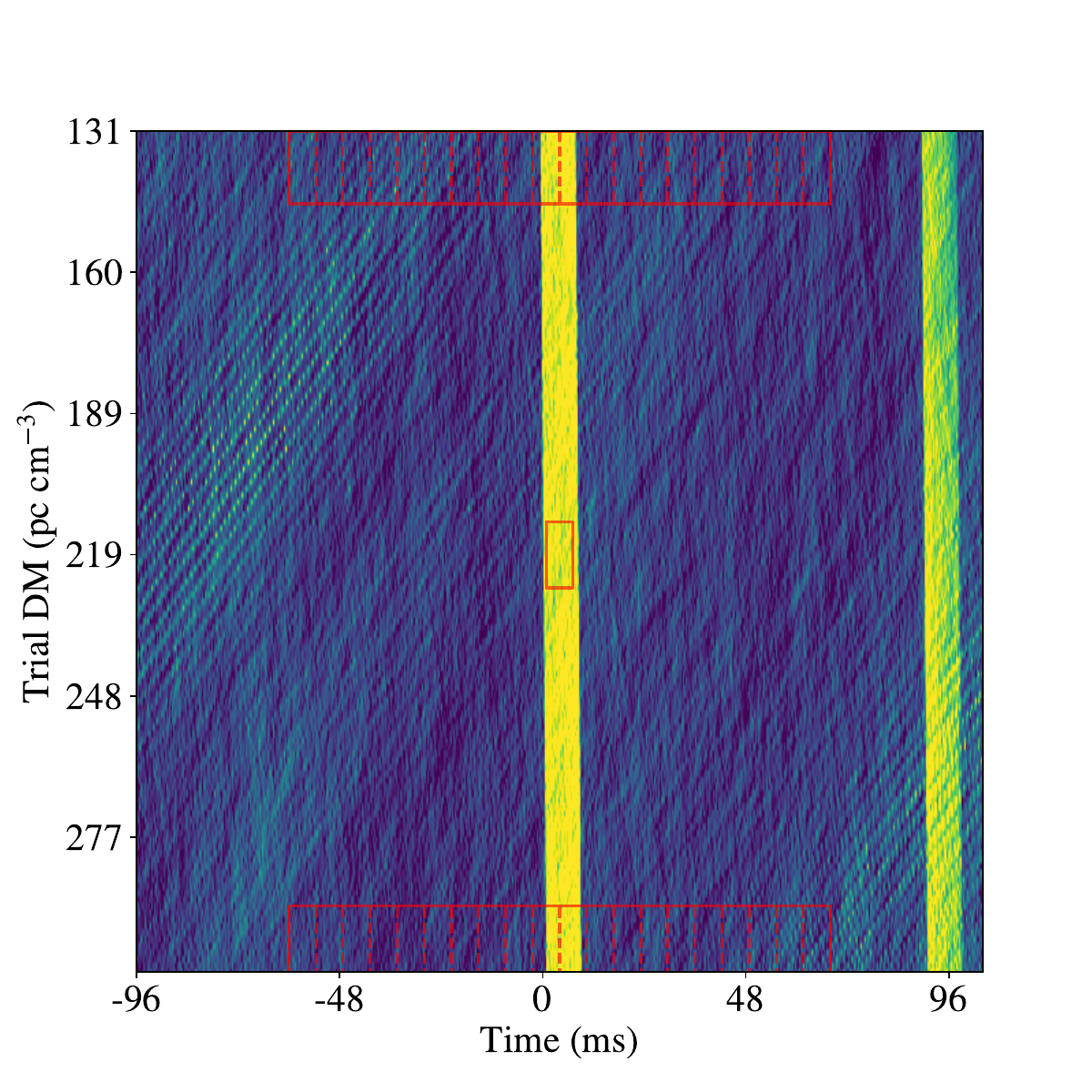}
    \caption{The DM-time spectrum of an RFI candidate from \FRB detected by the NRT.} \label{fig:rfi_catch}
  \end{subfigure}%
  \hspace*{\fill}   
  \begin{subfigure}{0.48\textwidth}
    \includegraphics[width=\linewidth]{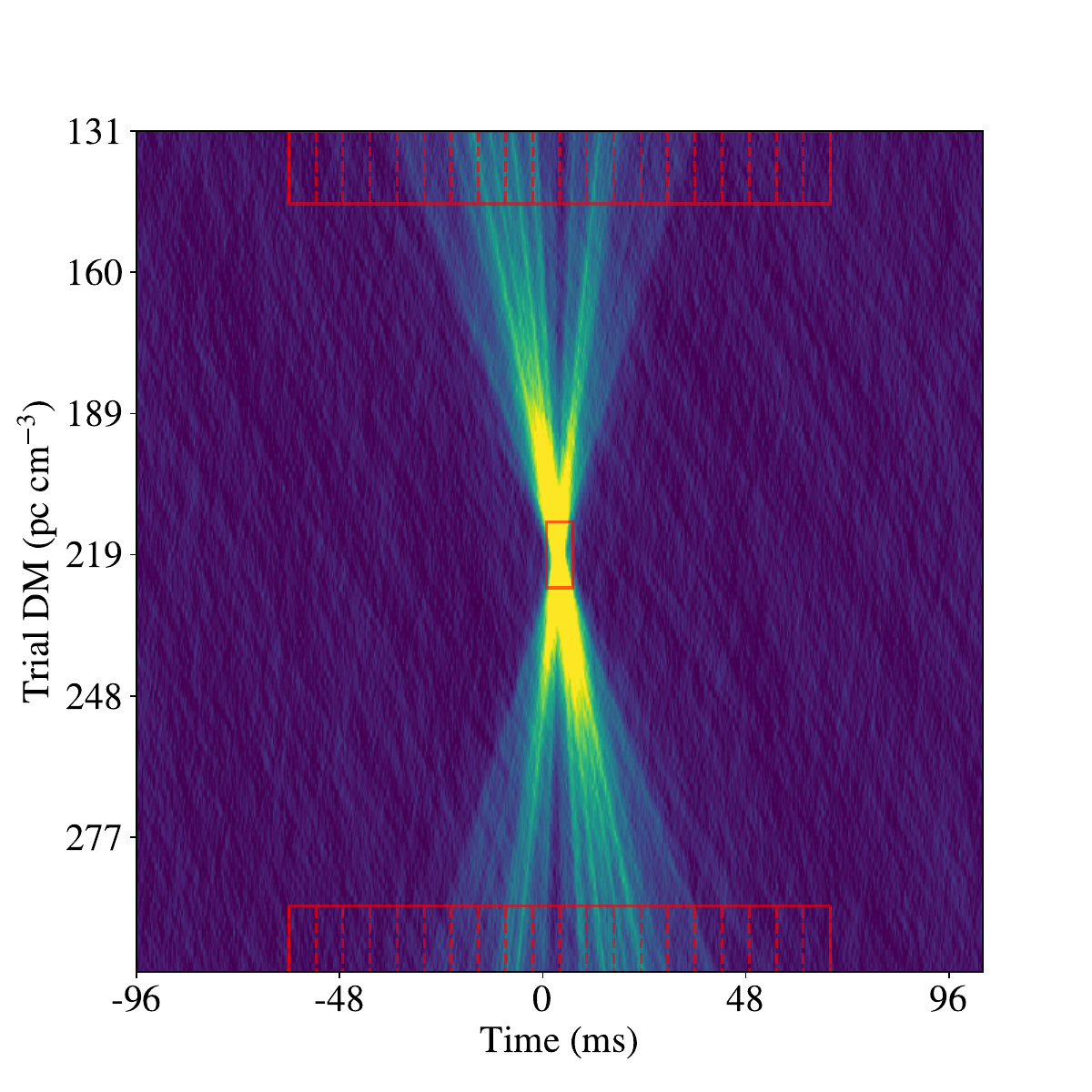}
    \caption{The DM-time spectrum for burst B049.\newline } \label{fig:frb_catch}
  \end{subfigure}%
\caption{Frequency-integrated brightness as a function of DM and time (`DM-time spectrum') are shown for an RFI candidate and for burst B049. In the center of the plot, a box is placed at \texttt{Heimdall}'s estimated DM value. Additionally, at the top and bottom of the plot, a set of boxes is positioned at the trial DM values where the burst would be smeared in time by a factor of ten. The dotted lines indicate the width of the region for which the intensity is calculated. The intensity of the RFI candidate does not decrease in the top and bottom regions compared to the central box. In contrast, for the FRB candidate, the intensity does decrease, signifying that this candidate is indeed an FRB.}
\label{fig:dmtrialburst}
\end{figure*}

An example is shown in Figures~\ref{fig:rfi_catch} and \ref{fig:frb_catch} for an RFI candidate and for burst B049, respectively. For the RFI candidate, the intensity remains relatively consistent at both the top and bottom of the plot, whereas, for burst B049, there is a substantial decrease in intensity. The ratio between these intensities is a reliable indicator for distinguishing between FRBs and RFI, and we dub this the `bow tie intensity ratio'. For the RFI candidate, the bow tie intensity ratio will never exceed 1.0. However, for burst B049, as the intensity decreases away from the centre, the bow tie intensity ratio will consistently be greater than one.

\subsection{Classification of burst candidates}
To assess the validity of our algorithm, we must look at the sensitivity. In this context, `sensitivity' is a classification metric that represents the number of true positives divided by the number of true positives plus the number of false negatives: 

\begin{equation}
    \text{Sensitivity} = \frac{\text{TP}}{\text{TP}+\text{FN}}.
	\label{eq:sens}
\end{equation}
\noindent

This metric penalises false negative predictions, thus penalising the misclassification of real FRBs. A perfect classifier would achieve a sensitivity of 1.0, as it would not misidentify any FRBs. 

Sensitivity is unaffected by the number of false positives, so optimising for sensitivity often results in a large number of potential FRB candidates that require manual examination. For instance, an algorithm can achieve a sensitivity of 1.0 by classifying all potential candidates as FRBs. We therefore try to optimise the sensitivity for our classifier while keeping the number of false positives below a manageable threshold. We have placed this threshold at 15 per cent of the total number of candidates. 

\begin{figure}
    \centering
    \includegraphics[width=1\columnwidth]{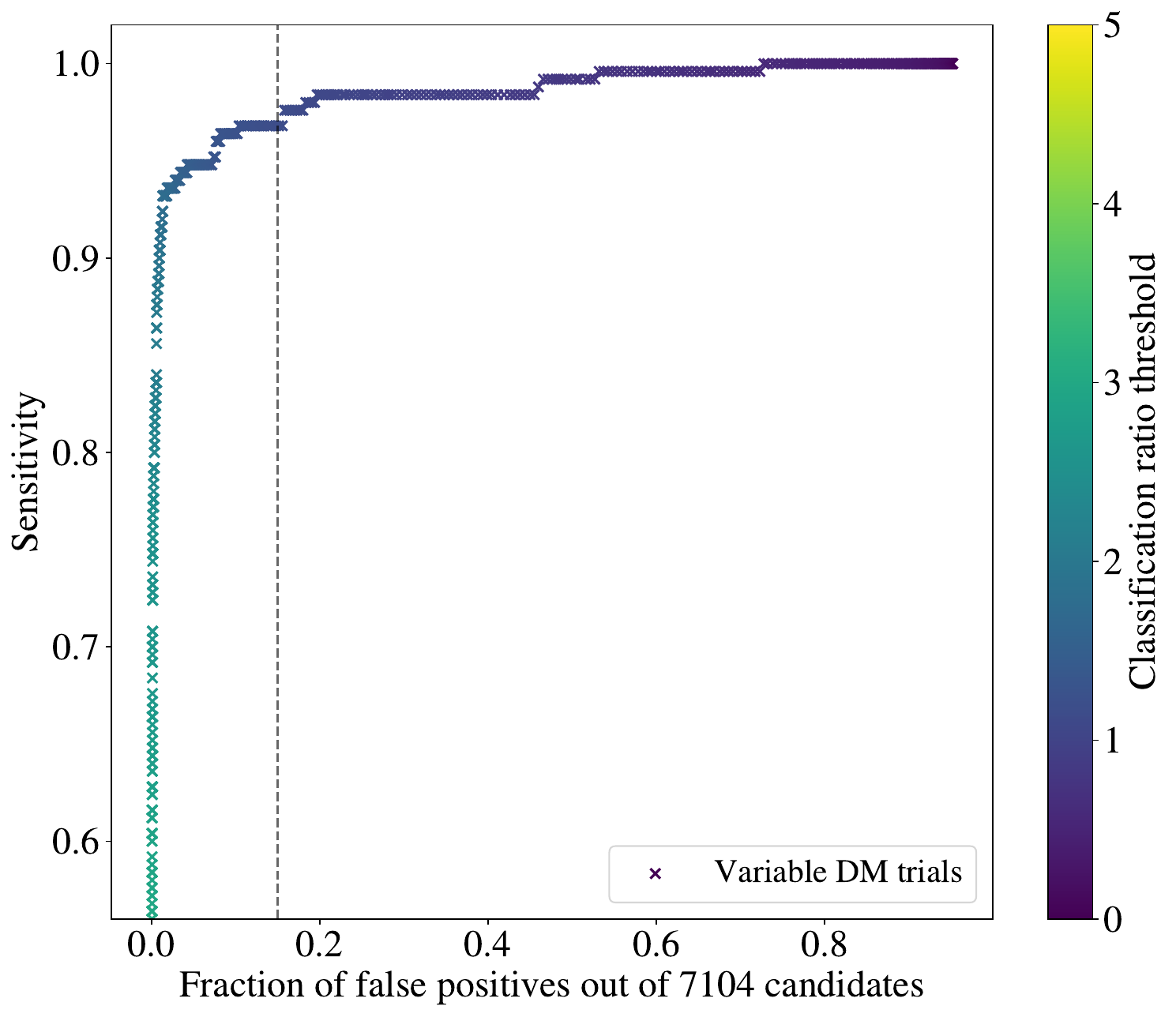}
    \caption{The sensitivity of the bow tie classifier (at different bow tie ratio thresholds) as a function of the percentage of false positives. The colour of the crosses indicates at what value the classification threshold has been placed. The dashed grey line representes the maximum fractional false positive limit, which has been set at 15 per cent of the total number of candidates.}
    \label{fig:sens_vs_fp_only_bowtie}
\end{figure}

\texttt{CATCH} classifies candidates based on the ratio between the average intensity in the central box and the highest average intensity in one of the boxes at the trial DM value that would smear the burst width by a factor of ten. To find this ratio, we have trained the algorithm using 7104 manually inspected burst candidates from eleven randomly selected datasets from \FRB. \texttt{CATCH} classified these candidates based on whether their bow tie intensity ratios exceeded a specific threshold value. By analysing \texttt{CATCH}'s performance at varying threshold levels, we can find the threshold at which the algorithm optimally performs. Figure~\ref{fig:sens_vs_fp_only_bowtie} shows the sensitivity of this bow tie method plotted against the number of false positives, where the colour of the crosses indicates the thresholds. The best performance is achieved at a treshold level of 1.24. \\

\begin{figure}
    \centering
    \includegraphics[width=1\columnwidth]{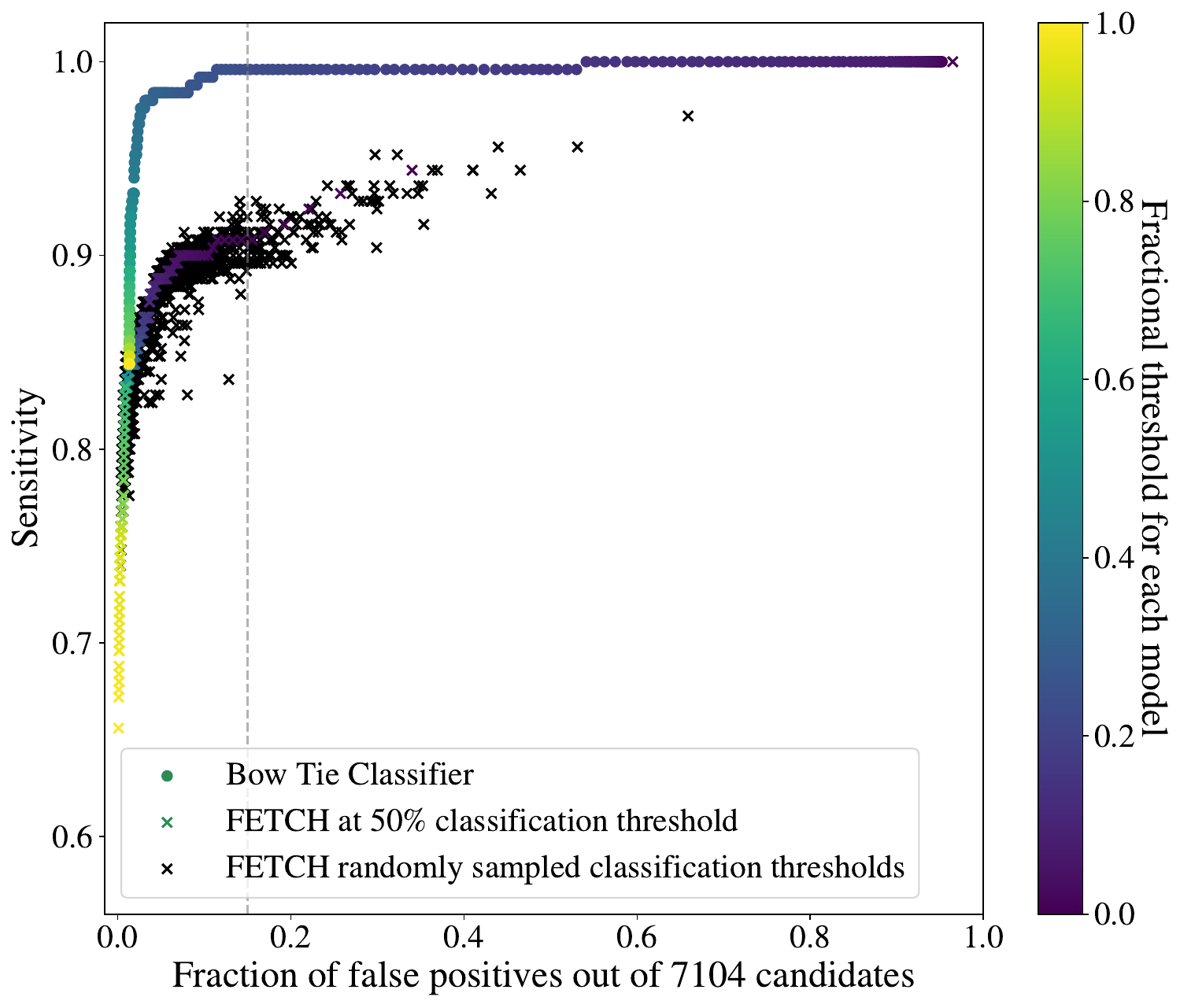}
    \caption{The sensitivity of \texttt{CATCH} and \texttt{FETCH} plotted against the percentage of false positives. \texttt{FETCH} has been plotted as crosses, including the static classification threshold across all models and randomly sampled thresholds. \texttt{CATCH} has been plotted as circles. The dashed grey line marks the 15 per cent fractional limit on the total number of candidates. At the dashed grey line limit, \texttt{CATCH} vastly outperforms \texttt{FETCH} in terms of sensitivity.}
    \label{fig:sens_vs_fp_fetch_vs_bowtie}
\end{figure}

Machine learning algorithms, like the conventionally used \texttt{FETCH}, demonstrate a higher level of robustness when confronted with singular noisy frequency channels in the DM-time spectrum, but have difficulty with large RFI structures. To create a classifier that achieves the highest sensitivity possible, \texttt{CATCH} includes the candidates that \texttt{FETCH} identifies as FRBs with a probability higher than 50 per cent, all the candidates which have a S/N ratio above 200, and all candidates that are selected by the bow tie method. The result is a significant increase in sensitivity, without exceeding the 15 per cent limit for false positives. Figure~\ref{fig:sens_vs_fp_fetch_vs_bowtie} shows the sensitivity of \texttt{CATCH} plotted against the number of false positives using varied classification ratios, like in Figure~\ref{fig:sens_vs_fp_only_bowtie}. The sensitivity of \texttt{FETCH} has been plotted for comparison, visualised with crosses, which includes the static classification threshold across all models and randomly sampled classification thresholds for each model separately. At the aforementioned 15 per cent maximum false positive threshold, \texttt{CATCH} attains a sensitivity of 99.62 per cent on the test dataset, surpassing the sensitivity of the conventionally used machine learning algorithm by $\sim$29 times.\\

\texttt{CATCH} has proven valuable, finding bursts in the training data set with a sensitivity of 99.6 per cent, and an accuracy of 88.8 per cent, while achieving a false positive total below the 15 per cent imposed limit. In the 60.99 hours of data obtained from the hyperactive \FRB, \texttt{CATCH} successfully detected 696 FRBs, surpassing the number that the standard method would have found by 26 per cent. Although \texttt{CATCH} displays promise, it is pertinent to acknowledge the identified limitations and challenges. First, to achieve such high sensitivity, \texttt{CATCH} classifies a greater number of candidates as FRBs, leading to a higher number of false positives. This approach proves valuable when analysing data containing many FRBs, as it vastly reduces the likelihood of missing real FRBs. However, for real-time systems where the false positive rate \textit{must} be low, or when monitoring sources that are not hyperactive, this method is not preferable. Secondly, when evaluating the types of bursts that conventional classification algorithms (such as \texttt{FETCH}) misclassify but \texttt{CATCH} correctly identifies, we find that lower S/N bursts are typically misclassified. However, the conventional classification algorithm also misclassified several brighter bursts (S/N $\sim$ 20). A comprehensive discussion of these issues has been undertaken by \citet[][]{DavidThesis}, and for brevity, we will refrain from further delving into the details here.

\section{Determining the dispersion measure uncertainty}
\label{ap:dm}

\begin{figure}
    \centering
    \includegraphics[width=1\linewidth]{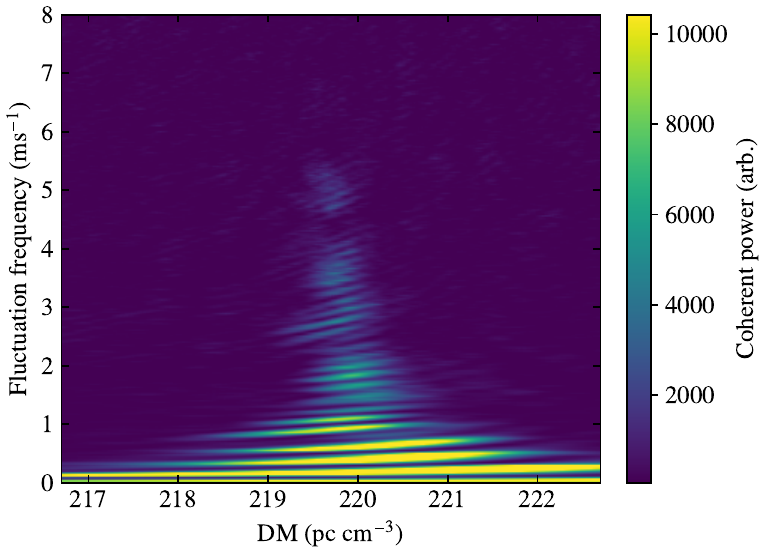}
    \caption{The coherent power of burst B586, plotted for a range of fluctuation frequencies against a range of DM values. The higher-intensity areas indicate vertically aligned structures across the frequency band. At the lower-end of the fluctuation frequencies, the burst exhibits vertical structures across the entire range of DM values. However, at higher fluctuation frequencies, only the small-scale structures align, specifically when the DM is 219.717\,pc\,cm$^{-3}$, indicative of the true DM of the burst. There is no apparent coherent power above fluctuation frequencies of 5.5\,ms$^{-1}$, which is the maximum fluctuation frequency of this burst.}
    \label{fig:fluctfreq}
\end{figure}

We estimated the error associated with the DM by calculating the $\delta$DM that corresponds to the inverse of the highest fluctuation frequency at which the burst still exhibits significant intensity in the power spectrum. Figure \ref{fig:fluctfreq} shows the coherent power of burst B586, plotted for various fluctuation frequencies against a range of DM values. At the DM of 219.717\,pc\,cm$^{-3}$, we can see that the coherent power extends to fluctuation frequencies of $\sim$5.5\,ms$^{-1}$. The inverted fluctuation frequency indicates the shortest intensity fluctuations of the burst (182\,$\upmu$s). Subsequently, we can calculate the number of DM units required to induce a variation on this timescale, corresponding to the error on the DM. 

\section{Wait-times of individual observations} \label{sec:waittimessingle}
\begin{figure}
    \centering
    \includegraphics[width=1\linewidth]{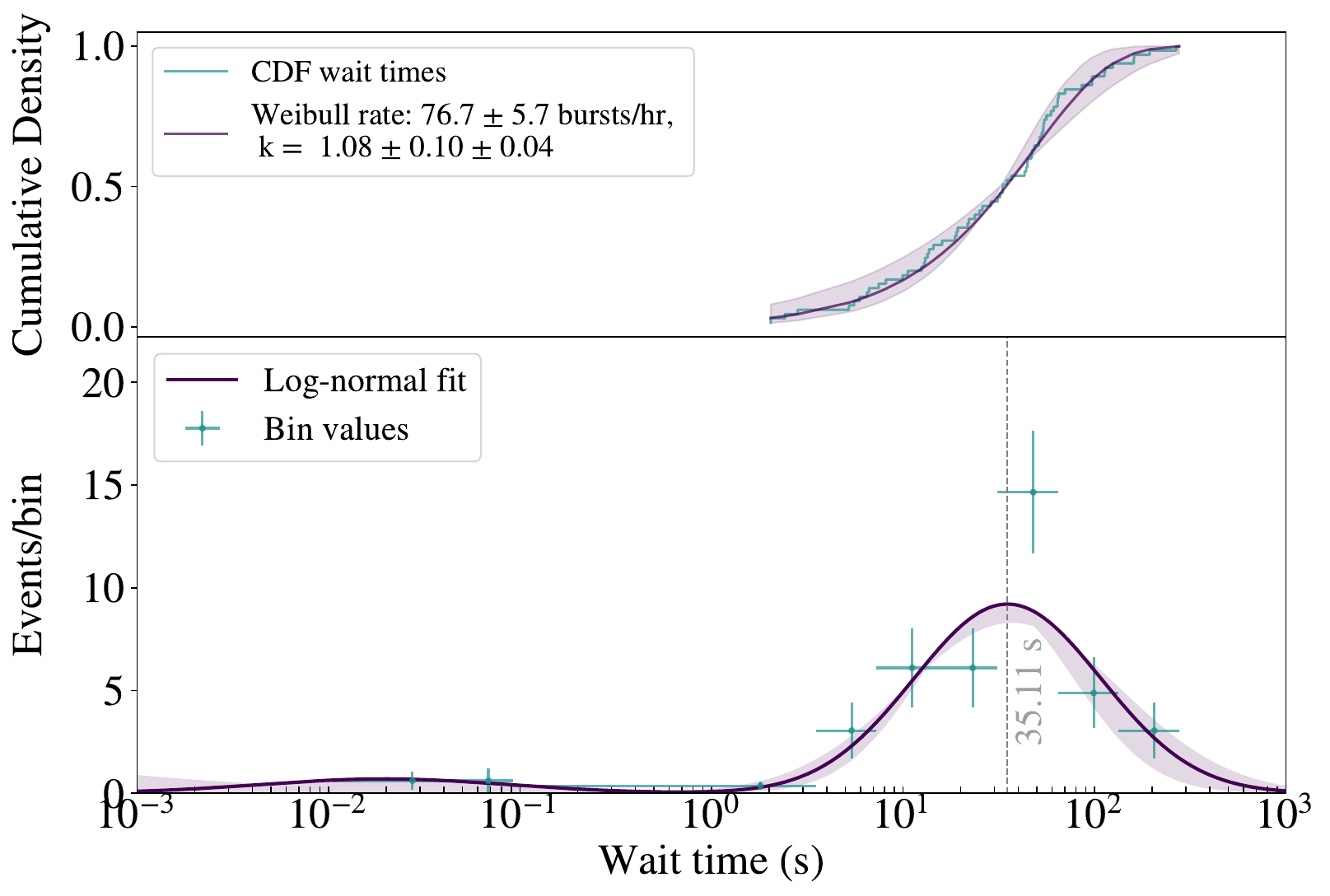}
    \caption{The distribution of the wait-times from the observation on MJD~59877, with a Poisson burst rate of $75^{+10}_{-9}$ bursts per hour. The cumulative wait-time distribution is shown in the top plot and fitted with a Weibull CDF. The shape parameter $k=1.08 \pm 0.10 \pm 0.04$ does not significantly deviate from $k=1$, which indicates the Poisson distribution can not be rejected as a potential fit to the data. Thus, for this observation, there is no significant evidence for clustering.}
    \label{fig:waittimes_59877}
\end{figure}

The wait-time distribution for the observation at MJD~59877 is plotted in Figure~\ref{fig:waittimes_59877}. During a period of 0.96\,h, a total of 72\,FRBs were detected, indicating a burst rate of $75^{+10}_{-9}$\,bursts per hour, where the errors are Poissonian. The best-fit Weibull CDF indicates a burst rate of $76.7 \pm 5.7$\,bursts per hour, which is consistent with the calculated burst rate. The shape parameter of the Weibull CDF is $1.08 \pm 0.10 \pm 0.04$, which is consistent with $k=1$, indicating that the Poisson distribution hypothesis can not be rejected. Thus, for this observation, there is no significant evidence for clustering.

\begin{figure}
    \centering
    \includegraphics[width=1\linewidth]{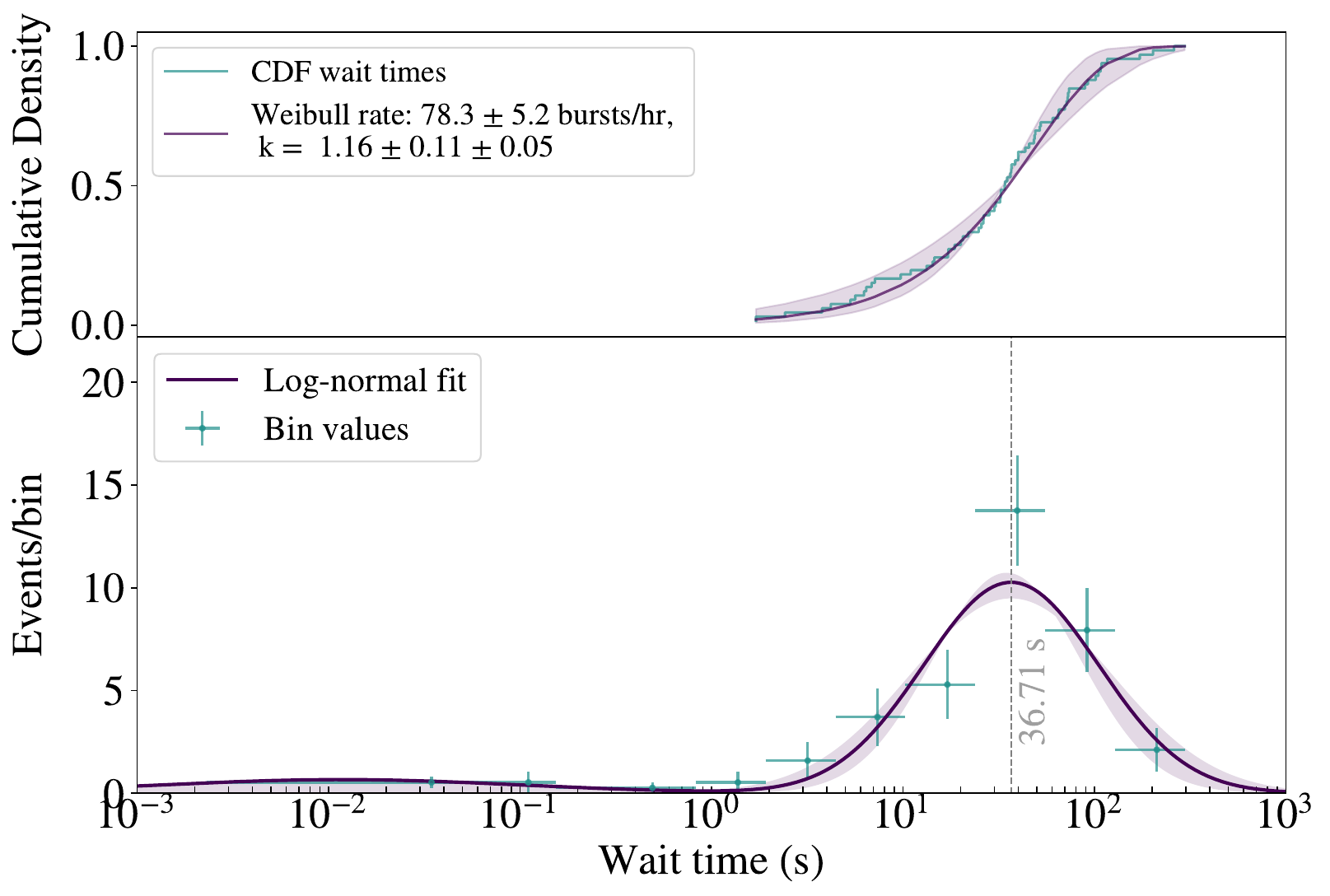}
    \caption{The distribution of the wait-times from the observation on MJD~59879, with a Poisson burst rate of $60^{+7}_{-7}$ bursts per hour. The cumulative wait-time distribution is shown in the top plot and fitted with a Weibull CDF. The shape parameter $k=1.16 \pm 0.11 \pm 0.05$ does not significantly deviate from $k=1$, which indicates the Poisson distribution can not be rejected as an accurate fit to the data. Thus, for this observation, there is no significant evidence for clustering.}
    \label{fig:waittimes_59879}
\end{figure}

The wait-times for the observation on MJD~59879 are shown in Figure~\ref{fig:waittimes_59879}. During a period of 1.37\,h, a total of 82\,FRBs were detected, giving rise to a Poisson burst rate of $60.06^{+7.35}_{-6.58}$ bursts per hour. The best-fit Weibull CDF indicates a burst rate of $78.3 \pm 5.2$\,bursts per hour, which is not consistent with the calculated burst rate. The shape parameter of the Weibull CDF is $1.16 \pm 0.11 \pm 0.05$, which is within errors consistent with $k=1$, again indicating that the Poisson distribution can not be rejected. Also for this observation, there is no significant evidence for clustering. 

\section{Observation Durations}
The 68 total observations, each lasting approximately 1 hour, are summarized in Table~\ref{tab:obs_log}.
\begin{table*}
\caption{\,~ NRT observation log for \FRB}
\label{tab:obs_log}
\begin{tabular}{c c c c @{\hskip 1.5in} c c c}
\hline \hline
Observation ID & Start (MJD)$^{\text{a}}$ & Duration (s) & & Observation ID & Start (MJD)$^{\text{a}}$ & Duration (s) \\ \hline
O1  & 59867.90178 & 1298.7 & & O35 & 59906.76284 & 2425.7 \\
O2  & 59869.87766 & 939.6  & & O36 & 59907.77528 & 2660.7 \\
O3  & 59870.86635 & 2004.7 & & O37 & 59909.75402 & 3952.2 \\
O4  & 59871.86137 & 3649.0 & & O38 & 59911.74708 & 4103.1 \\
O5  & 59873.85096 & 4138.0 & & O39 & 59912.73449 & 3417.0 \\
O6  & 59874.84975 & 2429.7 & & O40 & 59915.73991 & 3792.0 \\
O7  & 59875.85090 & 3578.9 & & O41 & 59917.73277 & 3980.9 \\
O8  & 59876.84022 & 2172.6 & & O42 & 59919.72279 & 4285.1 \\
O9  & 59877.84736 & 3451.9 & & O43 & 59921.72711 & 3451.9 \\
O10 & 59878.83732 & 2560.8 & & O44 & 59923.71440 & 4139.1 \\
O11 & 59879.82461 & 4915.2 & & O45 & 59925.70626 & 4285.0 \\
O12 & 59881.82641 & 4286.1 & & O46 & 59927.69795 & 4541.1 \\
O13 & 59882.82632 & 2562.8 & & O47 & 59931.68961 & 4340.2 \\
O14 & 59883.84423 & 2390.7 & & O48 & 59935.68565 & 3826.6 \\
O15 & 59884.82111 & 1884.6 & & O49 & 60007.49410 & 2948.7 \\
O16 & 59884.84769 & 1849.9 & & O50 & 60011.47434 & 2559.7 \\
O17 & 59885.81839 & 2559.8 & & O51 & 60012.47868 & 3451.9 \\
O18 & 59887.80276 & 3418.0 & & O52 & 60014.47088 & 3649.0 \\
O19 & 59888.81363 & 3792.0 & & O53 & 60016.46052 & 4138.1 \\
O20 & 59889.80460 & 2848.8 & & O54 & 60019.46660 & 3652.0 \\
O21 & 59890.80948 & 3648.1 & & O55 & 60020.44704 & 2846.8 \\
O22 & 59891.80041 & 2706.8 & & O56 & 60021.45762 & 3199.8 \\
O23 & 59892.80272 & 1748.7 & & O57 & 60023.45200 & 3182.9 \\
O24 & 59892.82765 & 1696.8 & & O58 & 60026.44155 & 3335.9 \\
O25 & 59893.79641 & 2560.8 & & O59 & 60030.41960 & 4284.0 \\
O26 & 59894.78365 & 4915.3 & & O60 & 60033.41924 & 3274.9 \\
O27 & 59896.78825 & 4136.1 & & O61 & 60036.41660 & 2678.8 \\
O28 & 59897.78272 & 2897.0 & & O62 & 60040.40335 & 2950.0 \\
O29 & 59898.78065 & 4272.2 & & O63 & 60042.40042 & 2671.4 \\
O30 & 59899.78016 & 2559.8 & & O64 & 60044.39492 & 2678.8 \\
O31 & 59900.77444 & 4336.1 & & O65 & 60047.38420 & 2950.0 \\
O32 & 59901.76453 & 3417.0 & & O66 & 60049.38117 & 3199.9 \\
O33 & 59903.76904 & 4178.0 & & O67 & 60051.37340 & 3314.9 \\
O34 & 59904.76995 & 2288.6 & & O68 & 60054.36522 & 3314.8 \\ \hline 
\hline
\multicolumn{1}{c}{$^{\text{a}}$ Topocentric at NRT.}\\
\end{tabular}
\end{table*}

\bsp	
\label{lastpage}
\end{document}